%% file: AGN_Paper_revisions_Jan2020.tex
\documentclass{aastex62}

\received{December 20, 2019}
\revised{February 5, 2020}
\accepted{February 21, 2020}
\accepted{February 21, 2020 in ApJ}

\graphicspath{{./}{figures/}}

\shorttitle{MACSJ1149 Variability Survey}
\shortauthors{Della Costa et al.}

\begin{document}

\title{Decadal Variability Survey in MACSJ1149}

\correspondingauthor{John Della Costa III}
\email{jmdc112596@ufl.edu}

\author[0000-0003-0928-2000]{John Della Costa III}
\affil{Department of Astronomy, University of Florida}

\author[0000-0002-2304-0908]{Vicki L. Sarajedini}
\affiliation{Department of Physics, Florida Atlantic University}

\author[0000-0002-7756-4440]{Louis-Gregory Strolger}
\affiliation{Space Telescope Science Institute}

\begin{abstract}

We present a long temporal baseline variability survey in the Frontier Field MACSJ1149. In this study, we identify active galactic nuclei (AGNs) and other transient sources via their variability using over a decade of {\sl Hubble Space Telescope} ({\sl HST}) images for thousands of galaxies in the cluster region and detect significant variability in galaxies extending down to an apparent nuclear magnitude of m$_{i}$ $<$ 26.5. Our analysis utilizes {\sl HST} images obtained in six different wavelengths from 435 nm to 1.6 microns and covers time scales 12 hours to 12 years apart. We find that $\sim$2\% of galaxies in these images are variable with 49 AGN candidates and 4 new supernovae candidates detected. Half of the variables are in the cluster and these are primarily elliptical galaxies displaying variability only in the near-infrared bands. About 20\% of the AGN candidates have morphologies and colors consistent with quasars, though most of the variables appear to be dominated by the host galaxy light. The structure function for these sources show a greater amplitude of variability at shorter wavelengths with slopes shallower than typical quasars. We also report a previously unknown Einstein cross identified in this field.

\end{abstract}

\keywords{Active galactic nuclei, Variable radiation sources, Hubble Space Telescope, Supernovae}

\section{Introduction} \label{sec:intro}

Active galactic nuclei (AGNs) are galaxies accreting significant amounts of material onto their central supermassive black holes (SMBHs). It is now believed that SMBHs exist in the centers of all galaxies with a significant bulge component \citep{kormendy95} and may have an important impact on the host galaxy's evolution. The observed relation between the mass of the SMBH and the bulge stellar mass \citep{ferrarese00,gebhardt00} suggests a coupled assembly history between the two. To understand the role played by AGNs in the evolution of galaxies, complete samples of AGNs in galaxy surveys spanning a range of redshifts, luminosities, and environments are required. 

AGNs can be identified using many techniques including UV/optical color selection \citep{mark67,sg83,richards02}, spectroscopic signatures of broad emission lines, or narrow emission lines with distinct line flux ratios \citep{baldwin81,vo87}, X-ray emission \citep{fabbiano92,alexander03}, infrared color selection \citep{lacy04,stern05}, radio emission (e.g. \citealt{sw80}), and variability (e.g., \citealt{hook94,macleod10}). Many of these techniques, however, produce samples biased against galaxies where the AGN light is much less than that of the host galaxy. Additionally, most surveys in the optical/UV are biased against obscured, dusty AGN/host galaxies. In order to identify representative samples of AGNs, several different techniques and selection methods should be used. While it may be found that AGNs selected using different techniques represent separate populations experiencing different phases in AGN/galaxy evolution (e.g., \citealt{hickox09}), a less biased sample of AGNs will help to make this picture clearer and allow us to better interpret the overall frequency of accreting SMBHs in the Universe and the role they play within their host galaxies.

 AGNs are known to vary on timescales of months to years in the optical, with 90\% to 100\% of AGNs identified via other means observed to vary over the course of several years (e.g., \citealt{kkc86,Schmidt10}). The mechanism behind the variability is still uncertain, though the leading explanation involves accretion disk instabilities \citep{pereyra06}, accretion disk surface temperature fluctuations \citep{kelly09}, or changes in the amount of infalling material onto the SMBH (\citealt{hopkins06,ruan14}). Regardless of the physics behind variability, AGNs are observed to display brightness variations of several percent over many years. There is additional evidence of an increase in variability amplitude for intrinsically fainter AGNs (\citealt{bershady98,vb04,unai14}), making variability a particularly effective means for identifying low-luminosity AGNs. Once identified, analysis of the light curves can be used to study accretion disk physics. The light curves appear to be well-represented by a damped random walk (DRW) or, more broadly, a CARMA model \citep{kelly14} and parameters that describe the variability, such as the slope of the structure function and characteristic timescale, may correlate with the physical causes behind the flux changes (e.g. \citealt{koz16}). The variability can also make use of reverberation mapping techniques to constrain accretion disk size when correlations are found in light curves covering a range of wavelengths (e.g. \citealt{peterson98, lira15}). Thus, variability is an important tool not only for identifying AGNs, but it can provide important clues as to the nature of the variability itself.

In this paper, we aim to identify AGNs using variability in the Hubble Frontier field MACSJ1149. We build on the results of previous successful surveys to identify AGN candidates in the nuclear regions of galaxies found in deep {\sl Hubble Space Telescope} ({\sl HST}) surveys (e.g., \citealt{villforth10, saraj11, pouliasis19}). The high-resolution {\sl HST} images allow us to obtain accurate photometry within small apertures, and thus probes lower AGN/host galaxy luminosity ratios than can be done using ground-based images. A significant fraction (50\% to 70\%) of sources identified using this technique are not identified using other methods such as X-ray emission or optical colors. These studies have shown that variability-selected samples vastly improve the AGN survey completeness as well as providing the light curves needed to analyze the variable nature of the AGN. The unique multi-wavelength, multi-epoch {\sl HST} observations of this cluster can also be used to explore accretion disk physics through analysis of the variability signature. Our survey for AGNs in a massive cluster also provides a contrasting environment to compare with field surveys and can be used to explore differences in the overall density of variability-selected AGN in different environments as well as their radial distribution within the cluster. 

In Section\,~\ref{sec:style} we describe the data selection, reduction, and photometric analysis of the sources we detected in MACSJ1149. In Section\,~\ref{sec:displaymath1} we describe the procedure for identifying significant variables in the cluster images. In Section\,~\ref{sec:displaymath2} we describe the characteristics of the variable AGN candidates. In Section\,~\ref{sec:displaymath6} we discuss new supernovae identified in the field. The variability analysis is discussed in Section\,~\ref{sec:displaymath7}. Finally, in Section\,~\ref{sec:highlight}, we summarize the basic conclusions of this study. Throughout the paper, we assume a flat, cosmological constant dominated cosmology with parameter values $\Omega_{\Lambda}$ = 0.7, $\Omega_{M}$ = 0.3, and H$_o$ = 70 km/s/Mpc. The data presented in this paper are in the AB photometric system.

\section{Data Selection and Photometric Reduction} \label{sec:style}

The Hubble Space Telescope Frontier Fields \citep{lotz17} consist of six deep fields centered on strong lensing galaxy clusters from \cite{abell89} and the MACS survey \citep{ebeling01}. This Director's Discretionary Time campaign probed deeper into the Universe than ever before with several hundred {\sl HST} orbits using the ACS/WFC and WFC3/IR cameras. The gravitational lensing provided by the clusters allows for some of the faintest sources ever to be observed. In one of the clusters, MACSJ1149, the multi-imaged lensed Refsdal supernova (SN) was detected in {\sl HST} images taken in November 2014 \citep{kelly15}. These data revealed four resolved images of a background SN in an Einstein cross around an elliptical galaxy in the cluster. Since mass models of the cluster predicted another occurrence of the SN in a different lensed image of the same galaxy elsewhere in the cluster within a few years, many additional epochs of data were taken for MACSJ1149 up to and including the period of the next event, which was observed in December 2015 \citep{kelly16}. The result was extreme temporal coverage of the cluster, more than 70 epochs in some of the near-infrared (NIR) images. These data also allowed for the detection of the highest redshift individual star, Lensed Star 1, a B-type star magnified by a factor of $\sim$2000 and observed at a redshift of 1.5 in the same host galaxy as the Refsdal SN \citep{kelly18}. In addition, MACSJ1149 was one of the CLASH survey clusters \citep{postman12} which obtained several epochs of {\sl HST} images in 2010 through 2013.

\begin{figure}
\centering
\includegraphics[width=\textwidth]{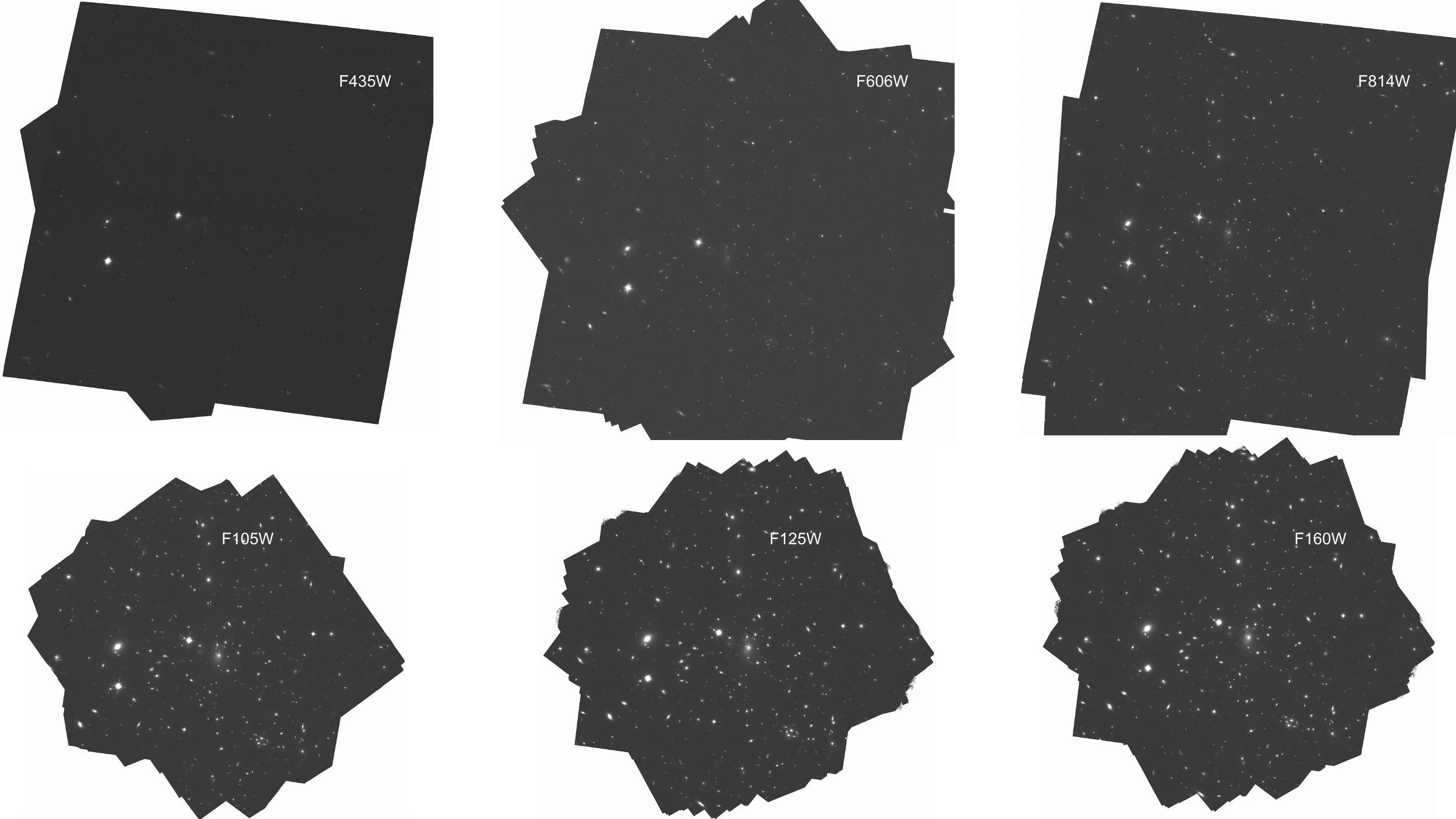}
\caption{Composite images of MACSJ1149 used in this survey. The top three images are the optical data that utilize {\sl HST}'s ACS Wide Field Camera, while the bottom images are the infrared data obtained with WFC3/IR. Each image is a stack of all epochs used in this study for that wavelength.}
\label{fig:image_panel}
\end{figure}
 
The wealth of high resolution imaging and temporal data in this cluster makes it an ideal candidate to search for AGN through their variability signatures. We chose to analyze the images of MACSJ1149 that cover the broadest range of wavelengths and time scales that have similar depth in each epoch to achieve our science goals. We searched the archive for images of MACSJ1149 that 1) had at least 1000 second exposure times and 2) had at least 10 epochs of data meeting the exposure time limit in a given {\sl HST} filter. This resulted in a total of 236 epochs in 6 different wavelengths (3 optical and 3 NIR) spanning more than 12 years at some wavelengths. Table 1 gives the total number of epochs analyzed, the range of time intervals sampled, and the average exposure time per epoch for each of the 6 filters analyzed in our study, including F435W (B-band), F606W (V-band), F814W (I-band), F105W (Y-band), F125W (J-band), and F160W (H-band).

\begin{deluxetable*}{ccccccc}[t!]
\label{tab:sample1}
\tablecaption{MACSJ1149 epoch data used in this study}
\tablewidth{0pt}
\tablehead{
\colhead{Filter} & \colhead{\# of epochs} & \colhead{$\Delta$t} & \colhead{Average Exposure Time (s)}}
\startdata
\input{tables/filter_data_table.tex}

\enddata
\end{deluxetable*}

Each epoch is composed of 3 or more separate FLC images (for ACS/WFC) or FLT images (for WFC3/IR) obtained from the Mikulski Archive for Space Telescopes (MAST). These images were aligned and combined using Astrodrizzle, which is provided through DrizzlePac and Pyraf \citep{gonzaga12}. We used the task Tweakreg in Astrodrizzle to align all of the frames to a common reference frame with the same pixel scale and central pixel location in the image. In this way, all sources identified in one epoch are at the same XY position in all epochs. All optical images (F435W, F606W and F814W) were aligned to the same reference image with 0.05$\arcsec$/pixel and image size $\sim$3.5$\arcmin$ on a side. All NIR epochs (F105W, F125W, and F160W) were aligned to the same reference image with 0.13$\arcsec$/pixel and image size $\sim$2.5$\arcmin$ on a side. Once each individual epoch was produced, we combined all of the epochs into a single image for each wavelength of data. This created a deep image aligned with all individual epochs that could be used for source detection. The total exposure times for the six wavelengths are 12.7 hours in B, 14.3 hours in V, 32.2 hours in I, 24.0 hours in Y, 20.4 hours in J, and 38.2 hours for H. Figure\,~\ref{fig:image_panel} shows the stacked images for each of the six wavelengths analyzed in this study.

To identify sources and perform aperture photometry, we used Source Extractor \citep{bertin96} on the stacked image. A low detection threshold was adopted to improve source completeness in the individual epochs. We identified $\sim$2000 sources in each of the optical images and $\sim$1000 in the IR images. We visually inspected the sources in each image to remove spurious detections around bright stars and near the edge of the stacked frame. While our aim is to monitor the nuclei of galaxies to search for AGN variability, we did not eliminate off-nuclear sources identified in irregular and/or disturbed galaxies. This was done to avoid removing any dual AGN while remaining sensitive to the possibility of detecting previously missed supernovae (SNe). We used multiple apertures for each source in the image, ranging from 3 through 20 pixels in diameter, corresponding to 0.15$\arcsec$ through 1$\arcsec$ in diameter, in the optical images. For the larger pixel scale of the WFC3/IR images, we used aperture sizes of 2 through 12 pixels in diameter, corresponding to 0.26$\arcsec$ to 1.56$\arcsec$ in diameter.  

We then used Source Extractor to obtain photometry for all sources in each individual epoch using the source list generated from the deeper image in each wavelength and with the same set of apertures. The photometric catalogs were then zero-point corrected to obtain the most precise relative photometry among the epochs. To determine the offsets, we computed the average magnitude difference for sources in each epoch against a common epoch in each wavelength. Small offsets of order $\pm$0.01 to $\pm$0.03 mag were applied to each epoch to bring them into relative photometric agreement with each other.

\section{Variability Selection} \label{sec:displaymath1}

Since our aim is to isolate the variability in galactic centers, we chose an aperture for the variability analysis that would include light from an unresolved source and exclude as much light as possible from the surrounding host galaxy. The FWHM of unresolved sources in the ACS/WFC images is about 2 pixels or 0.10$\arcsec$ and 0.23$\arcsec$ in the WFC3/IR images. Therefore, we chose to perform the variability analysis using apertures with diameters of 0.25$\arcsec$ for the optical and 0.52$\arcsec$ for the IR images. These apertures are about twice the FWHM of an unresolved source and were chosen to minimize light from the non-varying host galaxy while reducing variability noise due to changes in the PSF with time and location on the CCD. 

\begin{figure}
\centering
\includegraphics[width=\textwidth]{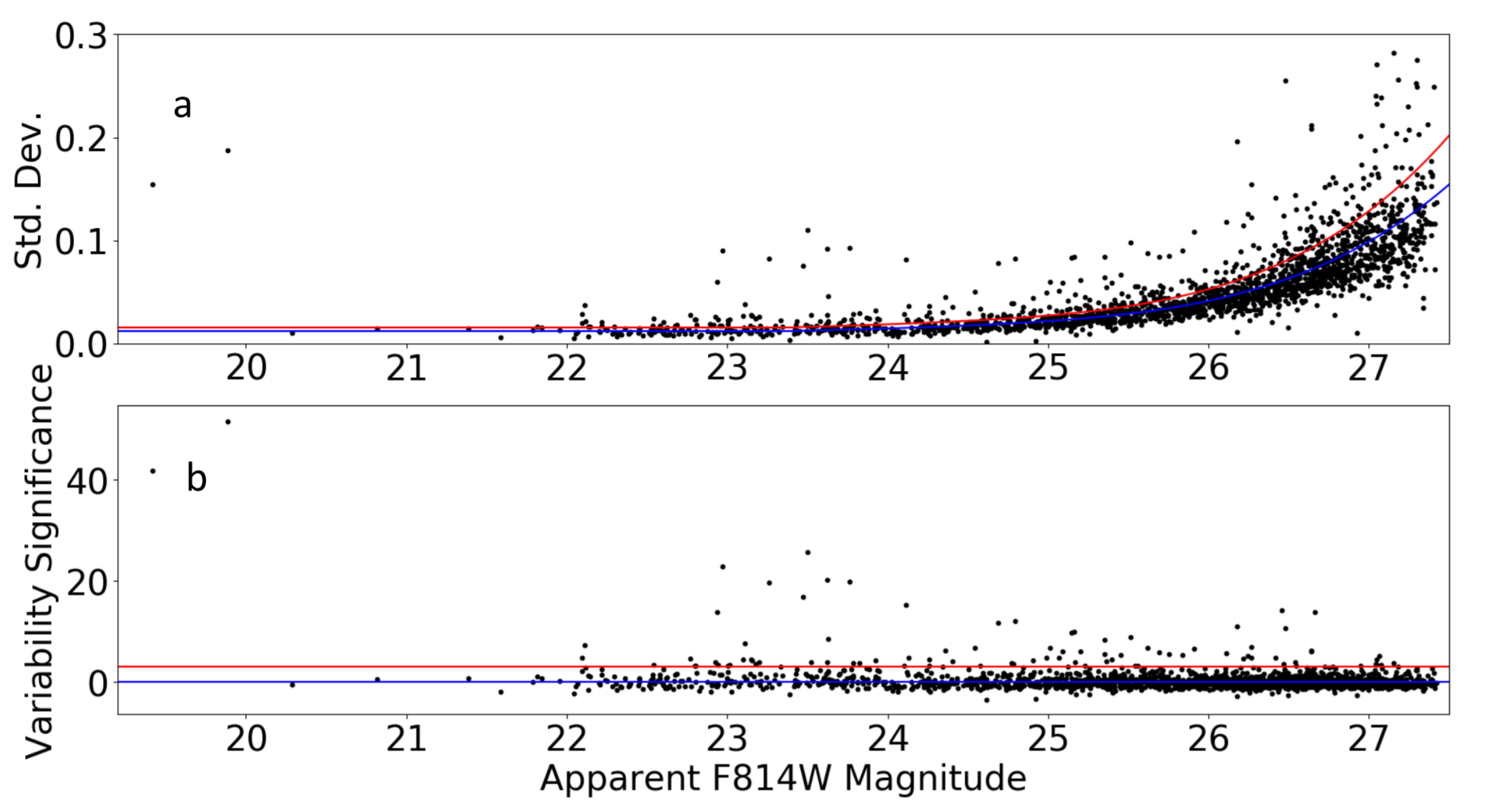}
\caption{a) The standard deviation of each galaxy vs. the apparent magnitude in the F814W band. The blue line represents a fit to the average standard deviation and the red line is the 1$\sigma$ spread of these values around the mean. b) The Variability Significance vs. the F814W magnitude. The Variability Significance is obtained by subtracting the mean (blue line in the upper panel) and dividing by the 1$\sigma$ spread (red line in the upper panel). The blue line indicates a variability significance of zero and the red line is 3$\sigma$.}
\label{fig:banana_curve}
\end{figure}

To quantify variability, we computed the standard deviation of the aperture photometry magnitudes for all sources measurable in at least three epochs for a given filter. We apply a small number statistics correction factor to our computed standard deviations for sources with fewer than 25 epochs of data. This factor accounts for the fact that, on average, a small data set will have a measured standard deviation smaller than its actual value. The multiplicative factor ranges from 1.12 for 3 epochs to 1.01 for 25 epochs. Figure\,~\ref{fig:banana_curve}a shows the corrected standard deviation value for all sources in the F814W band as a function of the average magnitude. The blue line in Figure\,~\ref{fig:banana_curve}a is the mean value of the standard deviations for sources of a given magnitude and the red line represents the 1-sigma spread of standard deviation values around the mean. Since the vast majority of sources are not expected to have intrinsic variations, this approach empirically quantifies the photometric noise as a function of source brightness. We then use the quantified noise to determine which sources have standard deviation values that are statistical outliers. More specifically, we subtract the mean value of the standard deviation at a given magnitude and divide by the 1$\sigma$ spread to determine the variability significance of each source. Figure\,~\ref{fig:banana_curve}b shows the variability significance as a function of magnitude for sources in the F814W band with the red line indicating the 3$\sigma$ threshold for significant variability. 

Because the photometric noise as a function of magnitude differs for sources detected in only a few epochs compared to those detected in many, this procedure was done separately for groups of galaxies detected in a small range of epochs. In particular, the median (blue line in Figure\,~\ref{fig:banana_curve}a) was found to be higher among galaxies that were measured in a larger number of epochs while the 1$\sigma$ spread (red line) was generally lower, as expected. By quantifying the variability threshold separately for sources found in a similar number of epochs, we were able to accurately remove biases in the variability significance while maximizing the number of galaxies available for the variability survey. We also investigated whether or not the photometric noise as a function of magnitude differed for extended and point-like sources. To do this, we computed the concentration index (CI), which is the difference between the aperture photometry in 10 and 5 pixel apertures. Point-like sources will have a small CI compared to extended sources. We separated the sources into point-like and extended subsamples based on the CI value and compared the standard deviation plot of Figure\,~\ref{fig:banana_curve}a made separately for each group. We found no significant difference in the distributions and therefore use the same thresholds for determining significant variability for point-like and extended sources.

\begin{deluxetable*}{crrccccccc}[t!]
\label{tab:sample2}
\tabletypesize{\footnotesize}
\tablecaption{Variable AGN candidates detected in MACSJ1149 including the variable's ID, right ascension (RA) and Declination (Dec), redshift, V, I, Y, J, and H band variability significance, and morphological classification. For the morphology class, 0 indicates elliptical, 1 is spiral, 2 is point-like, 3 is point-like with visible host, and 4 is irregular (includes tadpoles, mergers, etc.).}
\tablewidth{0pt}
\tablehead{
\colhead{ID} & \colhead{RA} & \colhead{Dec} & \colhead{Redshift} & \colhead{V} & \colhead{I} & \colhead{Y} &\colhead{J} &\colhead{H} &\colhead{Morphology} \\
\colhead{} & \colhead{(deg)} & \colhead{(deg)} & \colhead{$z$} & \colhead{} & \colhead{} & \colhead{} &\colhead{} &\colhead{} &\colhead{Class. ID}}
\startdata
\input{tables/variable_table.tex}
\enddata
\end{deluxetable*}


Once the variability threshold was quantified, we visually inspected each source above 3$\sigma$. This process revealed a number of spurious variables including 1) stars which appeared to be varying due to their proper motions across the frame, 2) sources where one or more cosmic rays affected the photometry in at least one epoch, and 3) sources that fell too close to the edge of the frame in one or more epoch. After removing these from our variability analysis, we are left with 1237 total objects in F435W, 1908 in F606W, 1815 in F814W, 746 in F105W, 736 in F125W, and 759 in F160W. 

\begin{figure}
\centering
\includegraphics[width=\textwidth]{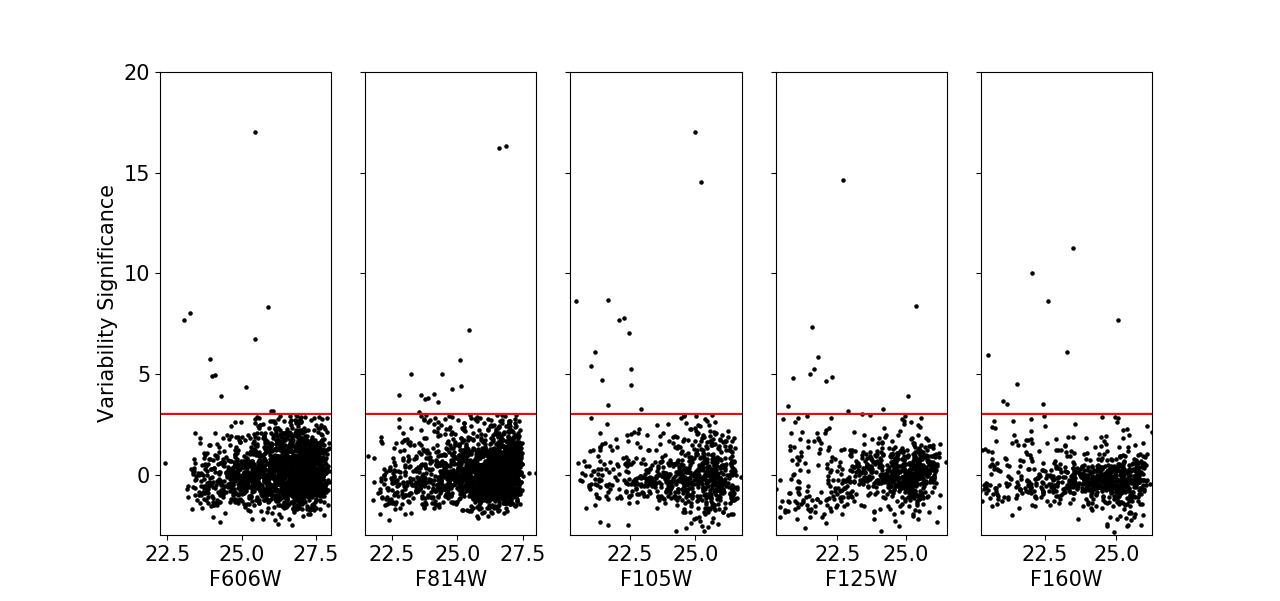}
\caption{Variability Significance vs. apparent nuclear magnitude for galaxies in the F606W, F814W, F105W, F125W, and F160W bands, respectively. The red line indicates the 3$\sigma$ threshold used to identify significant variables above the photometric noise.}
\label{fig:std_plots}
\end{figure}

Figure\,~\ref{fig:std_plots} shows the variability significance as a function of magnitude in each filter of our survey after removing spurious sources. The red line indicates the 3$\sigma$ significance above which we consider sources to be significant variables. Sources above this threshold are listed in Tables 2 (AGNs) and 3 (SNe). We identified 16 significantly varying galaxies in F814W and 14 in F606W. Unfortunately, due to the short temporal range covered by the B-band images and their relatively shallower depth, especially in the first two epochs, which provided the longer time baseline, we did not find any significant variables among the galaxies surveyed at this wavelength. Among the 16 variables identified in the I-band image, 4 are SNe candidates, and are discussed in detail in Section\,~\ref{sec:displaymath6}). Removing the likely SNe sources, we find a total of 22 AGN candidates among the optical bands with 4 of these sources (18\%) appearing as significant variables in both the V and I bands.

In the NIR images, we found 15 variables in F105W, 14 variables in F125W and 12 in F160W. If we remove the SN and SN candidates, the number of candidate AGNs are 14, 13, and 10, respectively. Five of the galaxies are variable in 2 bands and 1 is variable in all 3 NIR bands. Among the NIR variables, we find a total of 32 AGN candidates with 6 being variable in more than one of the NIR bands (19\%). Overall, we found 49 different variable galaxies that are likely to be AGN and 5 variables which appear to be SNe based on their lightcurves (one previously detected and 4 new; see Section\,~\ref{sec:displaymath6}). For the AGN candidates, 10 of the 49 are variable in at least two wavelengths (20\%), with one of these (\#6) displaying significant variability in 3 bands and another (\#8) displaying variability in all 5 bands.

To quantify the number of statistical outliers expected in our study, we examined the distribution of variability significance values for all sources in each filter. We find that the distribution can be modeled as Gaussian to $\sim$3$\sigma$ where it transitions to a shallower tail extending to high significance values. This distribution is expected when combining a normal, non-varying population with a distribution of true variables. The number of outliers expected beyond 3$\sigma$ for the normal distribution in each band is 0.27\% of the total sources, resulting in 5.2, 4.9, 2.0, 2.0, and 2.0 outliers in the F606W, F814W, F105W, F125W and F160W bands, respectively.  However, 80\% of these should have significance values less than 3.5$\sigma$ in a normal distribution. For example, of the $\sim$5 outliers expected in the F606W band, $\sim$4 should have values between 3 and 3.5$\sigma$ in a normal distribution with one having a value greater than 3.5$\sigma$.  The observed number of variables between 3 and 3.5$\sigma$ is just 2 with the remaining 12 sources having values greater than 3.5$\sigma$. Therefore, based on Gaussian statistics coupled with the true distribution of significance values, we estimate the number of statistical outliers to be no more than $\sim$3 (2 between 3 and 3.5$\sigma$, 1 greater than 3.5$\sigma$) in the F606W band. Using similar reasoning for the remaining bands results in $\sim$2, $\sim$2, $\sim$2, and $\sim$1 statistical outlier in the F814W, F105W, F125W and F160W bands, respectively.


\section{Variability Selected AGN Candidates in the MACSJ1149 Field} \label{sec:displaymath2}

In the five bands investigated in this study, we find that between $\sim$0.7\% and $\sim$1.9\% of galaxies in MACSJ1149 are significant variables likely to be AGN. If we account for the statistical outliers discussed in the previous section, the percentage of variable AGN is 0.6\% to 1.6\% of all sources. However, if we combine AGN variables detected in any of the five bands, the overall fraction is $\sim$2\% of the surveyed galaxies in the MACSJ1149 field.

Previous AGN variability surveys using {\sl HST} images find differing fractions of varying sources among the surveyed galaxies. A V-band survey of the GOODS North and South fields found that 2\% of all galaxies were significantly variables over a 6-month time interval \citep{saraj11}. Another study in this region conducted by \cite{villforth10} using the z-band ACS images identified a similar fraction of variables over the same time interval. The overlap between the V and z-band selected variables was 27\%. This is consistent with the level of overlap we find for variables selected at different wavelengths in this survey. More recently, \cite{pouliasis19} assembled images for the GOODS-S field over a 10 year period and found 0.5\% of sources in this z-band survey to be significant variables. Their sample had an 8\% overlap with the variables identified in \cite{villforth10} and a 17\% overlap with those identified in \cite{saraj11}. Differences in wavelength and the adopted source detection algorithm seem to be the primary reasons that variability surveys in the same field do not identify the same sources with a large degree of overlap. In general, however, the percentage of varying sources likely to be AGN found in similar {\sl HST} imaging surveys is consistent with the fraction we identify. Likewise, the degree of overlap we find for sources selected via variability at different wavelengths in our images is similar to that found among sources in the GOODS fields.

\begin{figure}
\centering
\includegraphics[width=\textwidth]{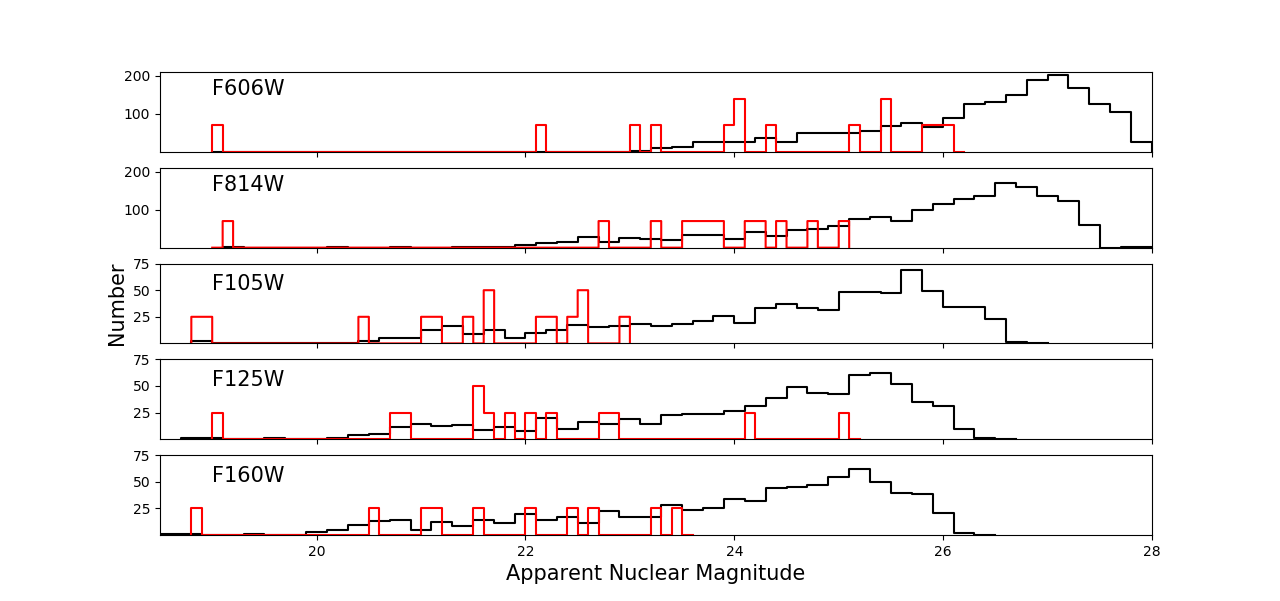}
\caption{The distribution of magnitudes (within 0.25$\arcsec$ diameter apertures for the optical bands and 0.52$\arcsec$ for the IR bands) for all galaxies (black line) and for the variable AGN candidates (red line). The AGN candidate histograms are multiplied by a factor of 70 for the optical and 25 for the IR bands for display purposes.}
\label{fig:histograms}
\end{figure}

The variable AGN candidates have brighter nuclear magnitudes than the overall distribution of galaxy nuclear magnitudes. Figure\,~\ref{fig:histograms} shows the distribution of magnitudes (within 0.25$\arcsec$ apertures for the optical bands and 0.52$\arcsec$ for the IR bands) for all galaxies (black line) and that for the variable AGN candidates (red line). The mean magnitude of the AGN are between 2 to 3 magnitudes brighter than the overall galaxy population in each waveband. The AGN may appear brighter for a number of reasons including the fact that brighter sources are more easily detected above the significant variability threshold, as shown in Figures\,~\ref{fig:banana_curve} and\,~\ref{fig:std_plots}. In addition to the AGN adding flux to the centers of galaxies, previous studies observed that AGN are generally found in brighter host galaxies (e.g. \citealt{saraj11}). The combination of these factors produces an overall brighter distribution of AGN nuclear magnitudes compared to the general galaxy population.

\subsection{Characteristics of the AGN Candidates} \label{sec:displaymath3}

We matched the galaxies analyzed in our survey with several other published catalogs for MACSJ1149 in order to obtain multi-wavelength galaxy photometry, photometric redshifts and, when available, spectroscopic redshifts. We first merged the F814W and F606W catalogs and included any sources that were also detected in the NIR images. We matched our sources with the Hubble Frontier Field DeepSpace catalog (HFFDS; \citealt{shipley18}) which includes 17 filters of {\sl HST} photometry for all galaxies in the region as well as spectroscopic redshifts. A total of 1981 galaxies were found to have matches within 0.2$\arcsec$ of a source in the HFFDS catalog. We also matched to the CLASH catalog \citep{postman12} which includes ACS/IR photometry and photometric redshifts determined using the Bayesian Photometric Redshifts technique \citep{coe06}. We found 2054 of our sources matched a source in the CLASH catalog to within 0.2$\arcsec$. Another large photometric catalog for the central part of the cluster was published by \cite{molino17} and includes 14 bands of photometry, photometric redshifts, stellar masses and B-band absolute magnitude estimates. We found 672 of our sources were matched to sources in this catalog. Finally, two additional catalogs of spectroscopic redshifts were matched to our sources; the CLASH VLT spectroscopic follow-up published in \cite{treu16} and \cite{grillo15}, and the \cite{ebeling14} follow-up with the Gemini and Keck II telescopes. We found 396 sources with a match in the CLASH VLT catalog and 127 in the \cite{ebeling14} redshift catalog. 

\begin{figure}
\centering
\includegraphics[width=\textwidth]{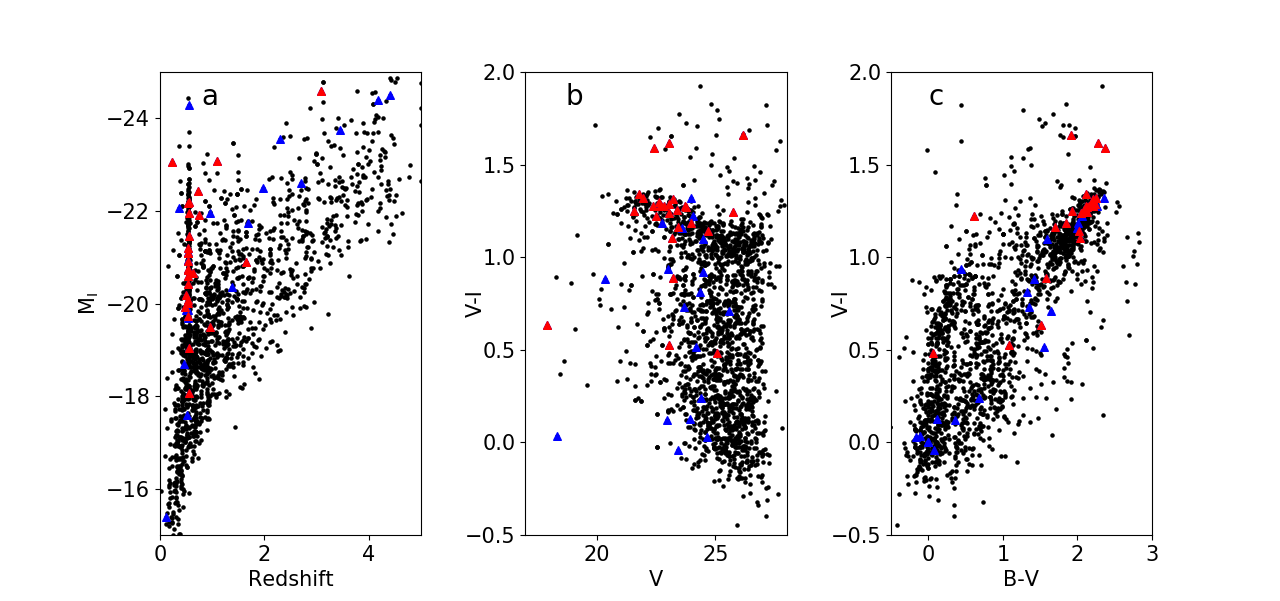}
\caption{Properties of galaxies and variability-selected AGN candidates in MACSJ1149. Black points are all sources in the survey with photometry and photometric/spectroscopic redshifts from catalogs described in the text. Blue triangles are variables detected in the optical bands (V and I) that may also display variability in one or more IR band. Red triangles are variables detected only in the IR bands. a) The redshift (photometric or spectroscopic) vs. absolute I magnitude. b) The V-band apparent magnitude vs. the V-I color. c) The B-V color vs. the V-I color.}
\label{fig:color_plots}
\end{figure}

Using these data we compare several properties of the overall galaxy population with those of the variable AGN detected in our survey. Figure\,~\ref{fig:color_plots}a shows the redshift vs. absolute I-band magnitude for the galaxies (black points) and variables (colored triangles) in our survey. The blue triangles are the variables identified in at least one optical band (V, I or both) and the red triangles are the variables found only in one or more of the 3 IR bands. Sixty percent of the variables have spectroscopic redshifts. The variables are at redshifts ranging from 0.1 to 4.2, with spectroscopic redshifts ranging from 0.4 to 1.7. About half (26 of 49) are at the cluster redshift of z = 0.55 with 4 at redshifts less than the cluster z and the other 19 at redshifts of 0.6 or higher. Those in the cluster are primarily elliptical in morphology (92\%). The variables make up 6.5\% of all galaxies at the cluster redshift, but only 1.3\% of galaxies in the foreground and background fields. A much higher fraction of IR-only variables are found in the cluster (70\%) compared to those variable in the optical (36\%). This is likely due to the higher fraction of elliptical galaxies in the cluster which are brighter in the IR bands. We find that the variables are found among the most luminous galaxies at all redshifts and particularly so among the higher redshift sources.

Figure\,~\ref{fig:color_plots}b shows the V magnitude vs. V-I color for the galaxies, using the same symbols as in Figure\,~\ref{fig:color_plots}a. The red sequence is clearly seen at V-I $\simeq$ 1.25. The majority of IR-only variables have colors consistent with elliptical galaxies, indicating that the overall color of the source is dominated by the host galaxy. Of the IR-only variables, 85\% have elliptical morphologies. The optical variables are primarily found in bluer galaxies whose colors resemble those of star-forming, spiral galaxies. Of those with optical variability, only 27\% have elliptical morphology with about half having point-like morphologies. The blue colors may also be influenced by the greater contribution of AGN light in these galaxies. Figure\,~\ref{fig:color_plots}c shows the observed B-V vs V-I color-color diagram for all sources with photometric measurements in all three filters. Again, the elliptical galaxies at the cluster redshift fall within a narrow range of colors at the red end of both color distributions and the IR-only variables are primarily located in this region of color-color space. However, there are a few exceptions where IR-only variables (red triangles) display bluer colors. For example, the source with red V-I color of 1.3 but with B-V $\simeq$ 0.6 is one of the Herschel detected sources (\#25) discussed below. The bright variable galaxy (\#8) has B-V and V-I colors close to zero. Several other optically detected variables have similar colors indicating that the AGNs dominate the host galaxy colors for these sources (about 20\% of the sources with BVI photometry). \cite{villforth12} showed that the majority of variability selected sources in the GOODS-S field were dominated by the host galaxy light. By fitting the SEDs of these sources with AGN and galaxy templates, they found that only 24\% of variability selected AGNs were dominated (more than 50\%) by the AGN component. This is consistent with the fraction of variables in our survey with optical colors like those of AGNs.

 \cite{rawle16} identified 39 sources in MACSJ1149 that have {\sl Herschel} IR emission. They combined multi-wavelength data for these galaxies to compute SEDs and derive various properties for the galaxies including star-formation rates and the AGN contribution. Of these 39 sources, we find 3 to be variable in at least one band of our survey. These are \#8 (variable in all five wavelengths), \#14 (variable in F606W and F105W) and \#25 (variable in F105W). Templates were fit to the SEDs that include the dust component, star forming component and an AGN component if necessary. They found that very few of the {\sl Herschel} detected sources required an AGN component of more than 10\% to accurately model the total IR emission (6 out of 263) and none of the three variables in our survey required the AGN template. However, if the variability indicates the presence of an AGN for our sources, the AGN can contribute to the SED in the mid-IR. This would result in an overestimation of the SFR. Indeed, object \#8 has a fairly high calculated SFR from the Rawle et al. SED fits (70 solar masses per year) which is likely an overestimation due to the contribution of the bright AGN in this galaxy.



\subsection{Cluster Radial distribution of AGN Variables} \label{sec:displaymath4}

\begin{figure}
\centering
\includegraphics[width=\textwidth]{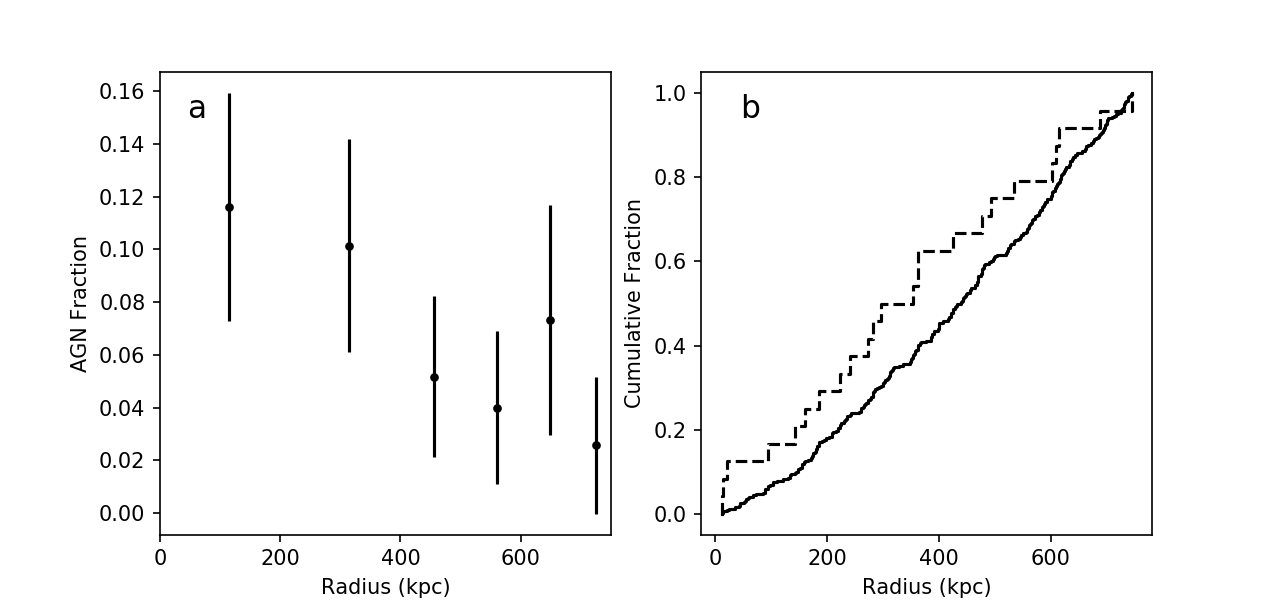}
\caption{a) The fraction of variable AGN candidates among galaxies in annuli of equal area as a function of cluster radius. b) The cumulative distribution of galaxies (solid line) and that of variable AGN candidates (dashed line) as a function of cluster radius.}
\label{fig:radial_dist}
\end{figure}

In Figure\,~\ref{fig:radial_dist}a, we show the radial distribution of variable AGN candidates in the cluster normalized by the galaxy number distribution. Radial bins of equal area are shown as a function of distance from the center of the cluster for sources with photometric or spectroscopic redshifts at the cluster redshift (0.5 $<$ z $<$ 0.6). This includes 372 galaxies and 24 variable AGN candidates. We find that the fraction of galaxies hosting a variable nucleus is greater in the center of the cluster ($\sim$10\%) than at a radius of 700 kpc ($\sim$3\%) which is about 30\% of the virial radius for MACSJ1149. We perform a Kolmogorov Smirnov test (KS test) to determine the probability that the AGN and galaxy radial distributions come from the same parent population. The test reveals that the AGN distribution is different from that of the normal galaxies with a statistically significant probability of 97.4\%. The cumulative distribution shown in Figure\,~\ref{fig:radial_dist}b confirms that the variability-selected AGN candidates are more centrally concentrated than the normal galaxies. 

A previous cluster AGN survey used optical variability, X-ray emission, and IR power-law selection for 12 clusters including MACSJ1149 \citep{klesman14}. They found that the X-ray selected AGNs were more centrally concentrated when combining the X-ray source radial positions for all clusters in their survey. The optically-varying sources showed a similar distribution to that of the normal galaxies, in contrast with our results. We find an overlap of only 25\% with the selected optical variables in their analysis. The smaller number of variables identified in their study (12) is likely a result of lower variability sensitivity since only two epochs of data in one wavelength were available for analysis of the cluster at that time. Our study is a more comprehensive survey for variability-selected AGNs in the optical and infrared, and therefore reflects a more complete sample of AGNs in the cluster. 

Our general result is consistent with that found for AGNs identified at various wavelengths in the central regions of galaxy clusters at a range of redshifts. \cite{galametz09} found that the X-ray and IR selected AGNs in their survey of hundreds of clusters in the NDWFS were enhanced in the central regions of the clusters and this enhancement increased with cluster redshift. Several other studies found similar results \citep{hart11, fassbender12, elhert13} with varying levels of central concentration for the AGN. \cite{mo18} identified AGNs using optical and IR colors as well as radio emission. They found that the radio and IR selected AGNs were more centrally enhanced than the optically selected AGNs, though all three selected samples identified more AGNs in the centers of clusters than in field surveys. These results support the theory that moderate ram pressure stripping \citep{marshall18} as well as galaxy harassment or tidal interactions among cluster galaxies can induce gas flow to the center of a galaxy, triggering an AGN. 

\subsection{Special Cases} \label{sec:displaymath5}

\begin{figure}
\centering
\includegraphics[width=\textwidth]{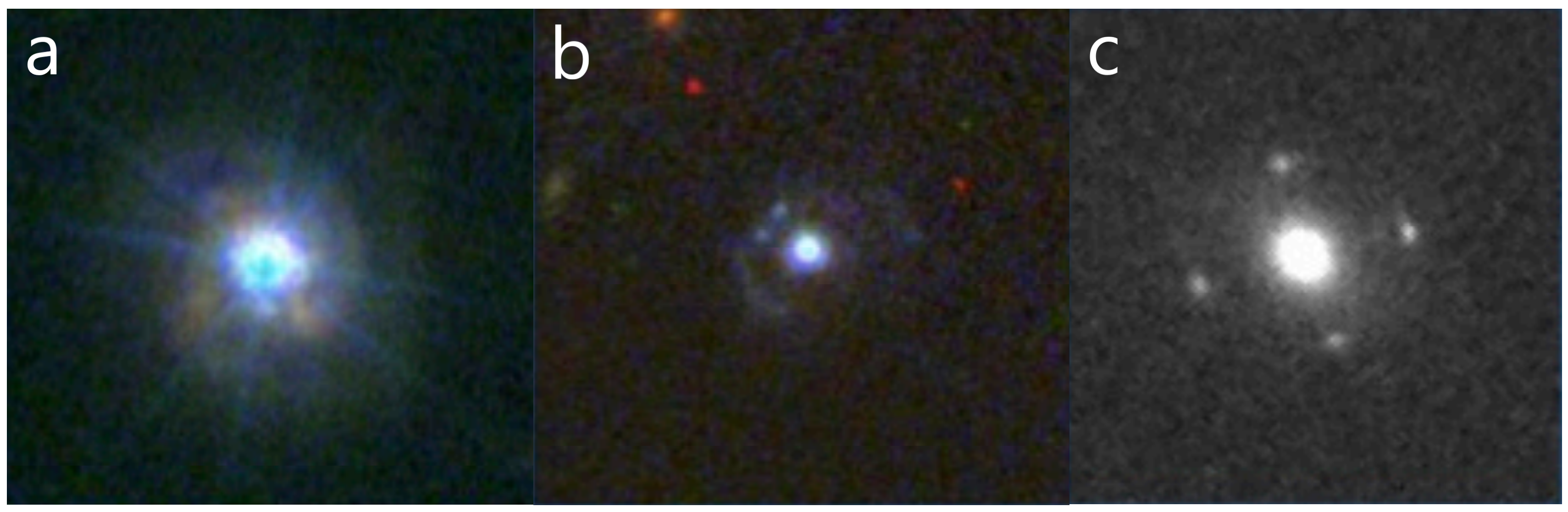}
\caption{a) Variable \#8 is a quasar with a visible host galaxy. It is the brightest AGN and the only one that is variable at all surveyed wavelengths in MACSJ1149. b) Variable \#13 is a point-like QSO that shows evidence of a host galaxy. c) The Einstein cross detected in one epoch of F606W.}
\label{fig:cool_images}
\end{figure}

Several sources stand out among the variable AGN in our survey. Variable \#8 is the only source with significant variability at all wavelengths analyzed in this study. This X-ray source has a bright, point-like nucleus within a face-on spiral host galaxy (Figure\,~\ref{fig:cool_images}a). The Structure Function (SF) for \#8 (Figure\,~\ref{fig:structure_functions}c) quantifies the amount of variability observed on different time scales and is further discussed in Section 5. The slope of the SF indicates that the variability is consistent with that observed for quasars \citep{koz16}. The absolute magnitude of the source is -24.23 as measured in the observed V-band images, which is similar to rest-frame B.


In January 2019, we obtained spectra for six of the variable sources detected in our survey using the OSIRIS spectrometer and multi-object mask mode on the Gran Telescopio Canarias (GTC). Of these, three were too faint to gain information from the spectra. One (\#14) showed narrow emission lines consistent with a star forming galaxy, and the other two (\#8 and \#13) showed broad emission lines consistent with AGNs. The spectrum of \#8 shows broad H$\gamma$ and some indications of weak [NeIII]. Unfortunately, the spectrum was impacted by light from an adjacent target on the mask, blocking the H$\beta$ line and making emission line measurements uncertain. Fortunately, this source was also observed as part of the Sloan Digital Sky Survey \citep{aguado19}. From the Sloan Digital Sky Survey (SDSS; DR15) spectrum, we can measure the broad H$\beta$ emission line to estimate the black hole mass for this AGN. Using equation 6 in \cite{VP06}, we find that this source has a black hole mass of $10^{8.4}$ $M_\odot$. Using the observed luminosity, we estimate the accretion rate of this AGN to be 44\% of the maximum Eddington accretion rate.

Another variable source of distinction is \#13 (Figure\,~\ref{fig:cool_images}b). The morphology of this variable AGN's host galaxy is unusual. It looks point-like, which means it is likely a quasar. However it does show some features that make it look disturbed, potentially from a recent merger. This could explain its high amplitude of variability. This source was only observed in the V-band images and has the greatest variability significance of any source in the survey. The color of this source is very blue and places it close to \#8 in the color-color diagram shown in Figure\,~\ref{fig:color_plots}c. The GTC spectrum reveals broad MgII emission at a redshift of 1.22. From the width of the MgII emission line, and using equation 7 from \cite{bahk19}, we estimate the black hole mass to be 10$^{8.7}$ $M_\odot$. Together with its absolute magnitude of -23.70, we find it is currently accreting at about 14\% of the maximum Eddington rate. 

Finally, we have identified a clear gravitationally lensed Einstein cross in the outskirts of one V-band image for this cluster (Figure\,~\ref{fig:cool_images}c), at RA=177.4029167 and Dec=+22.4366167. These optical structures are rare on the sky because they require a specific alignment between the lens and lensed source. The source lies more than 2$\arcmin$ from the cluster center and unfortunately falls in only one epoch of the V-band imaging data. For this reason we could not include it in our variability analysis, but the morphology allows us to roughly constrain the redshift of the lensed quasar. We have the redshift of the lensing galaxy from our GTC spectroscopy and measured it to be at the cluster redshift of z=0.54. Based on a comparison with a similar lens from \cite{bettoni19}, we estimate that the lensed system has a redshift between 1 and 3.
\section{Supernovae in the MACSJ1149 Field} \label{sec:displaymath6}

One of the primary scientific drivers motivating the large amount of temporal data in MACSJ1149 is the search for SNe. The well-known Refsdal SN was originally detected in late 2014 within a background galaxy lensed by the cluster \citep{kelly15}. Lensing models for the cluster predicted a later detection of the same event in another lensed image of the host galaxy which was observed in 2015 \citep{kelly16}. It is known as the first strongly lensed image of a SN resolved into multiple images. The SN is at a redshift of z = 1.49 and is being lensed by a cluster galaxy at z = 0.54. In addition to the Refsdal SN, there have been three more SNe found in the main MACSJ1149 field as part of the Hubble Frontier Field SN survey (S. Rodney PI; Strolger et al.,in prep, \citealt{kelly17}). The other three previously detected SNe are considered candidates since they were detected in only one epoch. 

\begin{figure}
\centering
\includegraphics[width=\textwidth]{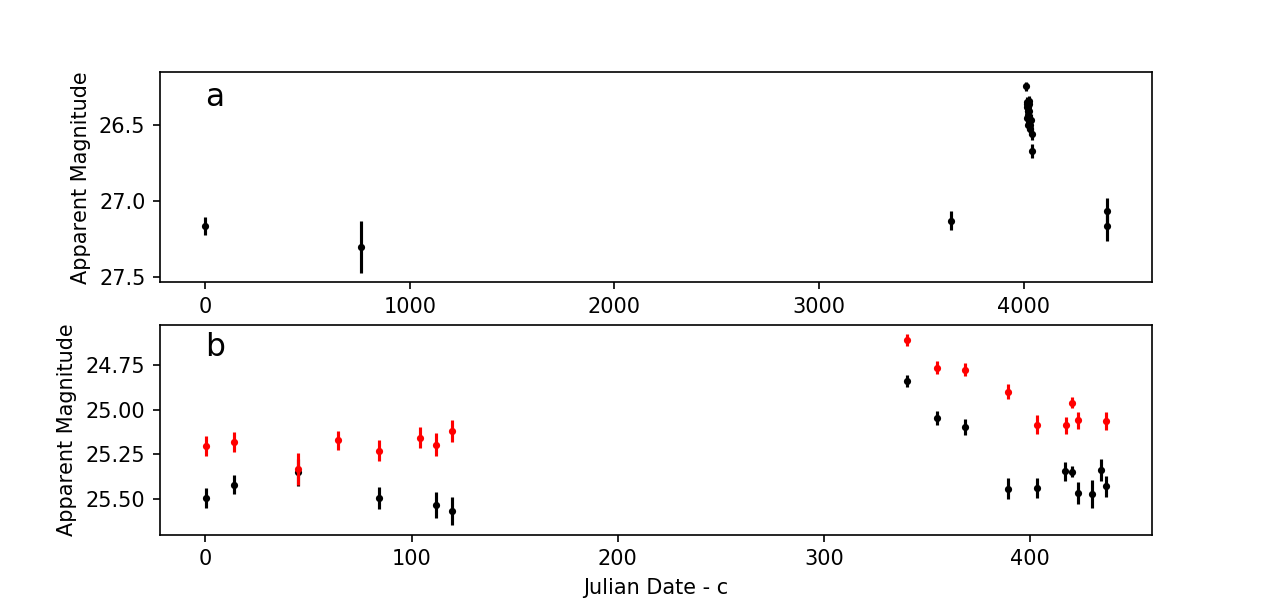}
\caption{Light curves for two newly detected supernovae. a) SN Dellanova with a maximum detected apparent magnitude at m$_{F814W}$ = 26.24 in April 2015. b) SN Saranova with a maximum detected apparent magnitude at m$_{F125W}$ = 24.84 (black) and m$_{F160W}$ = 24.61 (red) in February 2016. A constant $\it{c}$ of 53117.72 in the top panel and 57104.2 in the bottom panel is subtracted from the Julian Date.}
\label{fig:SN_lcs}
\end{figure}

In our study, we have found two new SNe with well-sampled lightcurves, and two more supernova candidates (Table 3). We also detected the Refsdal SN but do not include it in the table or this discussion since it was previously observed. Figures\,~\ref{fig:SN_lcs}a and\,~\ref{fig:SN_lcs}b show the lightcurves of the two supernovae with fading lightcurves detected in more than one epoch. The top panel (Figure\,~\ref{fig:SN_lcs}a) shows SN Dellanova, which appears in 25 epochs of the F814W data. This source falls outside of the FOV of the IR images and therefore no IR lightcurves are available. The host galaxy was detected in the V-band images, but the time intervals covered with the V-band photometry do not sample the period when the SN occurred. The median pre-SN apparent magnitude for this source is 27.20, measured as far back as 2004 at the SN location in the host galaxy. The maximum brightness detected occurs on April 18, 2015 at 26.24, an increase of almost one magnitude. We cannot definitively determine if this is the peak brightness and date since the previous measurement was taken one year earlier in April 2014 and the SN had not yet appeared at this time. The SN fades $\sim$0.3 mag over the next month of observations. At the next measurement taken one year later, the SN is back to its pre-SN magnitude of 27.16. The photo-z for this galaxy is 1.2 from the CLASH catalog. At this redshift, the peak luminosity of the SN is -18.38 at the observed wavelength, which is close to the B-band rest magnitude. This is almost one magnitude fainter than the typical peak brightness for a Type 1a SN of M${_B}=M{_V}$ = -19.3 \citep{hill00}, and would indicate that this source is either not Type 1a or that the observations do not sample the peak luminosity. 

The lower panel of Figure\,~\ref{fig:SN_lcs} shows the F125W (red) and F160W (black) lightcurves for SN Saranova. The temporal range of these lightcurves spans just over one year. The first several epochs before the appearance of the SN cover about 4 months from March 2015 to July 2015. The pre-SN brightness is 25.2 and 25.5 in the F125W and F160W bands, respectively. The next observations occur 7 months later, and the SN has appeared by February of 2016, having increased in brightness by $\sim$0.65 mag. It then fades over the next two months, back to the pre-SN brightness. The photo-z for this source is very uncertain, and the CLASH catalog gives several different redshifts ranging from 0.24 to 2.40 along the extent of the host galaxy. The morphology of the host galaxy indicates that it is lensed, so it is likely at a redshift greater than the cluster. At a redshift of 1, the peak brightness is close to that expected for Type 1a SN. Lacking better redshift information for this source, we cannot place firm limits on the absolute magnitude or nature of this SN event.

The other two SNe are candidates that appear in a single epoch. C1 is in 24 epochs in the I-band. The source has a magnitude of 24.95 in April 2004. The next epoch was obtained 2 years later. In this epoch, and all remaining epochs, the brightness falls and remains 0.2 magnitudes fainter than the first epoch. C2 is also in 24 epochs in the I-band. It begins at 25.5 in April 2004 and then rises to 25.18 in the next epoch obtained 2 years later. The next epoch was not obtained until 2014 and the source had returned to its original brightness 0.3 mag fainter than the peak. 

With the addition of the four SNe and SNe candidates presented here, the supernova rate for this field doubles. The additional sources detected in our study are likely a result of the different technique adopted for our analysis. The use of aperture photometry appears to be more sensitive to small changes within brighter host galaxies if the photometric noise can be accurately modeled from the non-varying galaxies. Most SNe surveys use difference imaging methods \citep{alard98}, which is more sensitive to SNe that outshine the host galaxy, and in some cases, where the host galaxy cannot be seen at all. Additionally, some of these new sources may be outside the area included in the SN survey or beyond the temporal range of the CLASH and Hubble Frontier Field images.

\begin{deluxetable*}{crrccccccc}[t!]
\label{tab:sample3}
\tablecaption{New SNe detected in MACSJ1149 field. } 
\tablewidth{0pt}
\tablehead{
\colhead{ID} & \colhead{RA} & \colhead{Dec} & \colhead{Band} & \colhead{Peak Date} & \colhead{Peak Apparent Brightness} & \colhead{Number of Epochs}\\
\colhead{Name} & \colhead{(deg)} & \colhead{(deg)} & \colhead{Filter} & \colhead{Month/Year} & \colhead{(mag)} & \colhead{}}
\startdata
\input{tables/sn_table.tex}

\enddata

\end{deluxetable*}




\newpage

\section{Variability Analysis and Structure Functions} \label{sec:displaymath7}

The variability of AGN can be described as aperiodic and appears to be due to a single stochastic process, or a combination of stochastic processes \citep{kelly11}. To quantify the behavior of the variability for the sources detected in our survey, we compute the Structure Functions for all sources detected in each of the 5 bands. The Structure Function is a measure of the rms variability as a function of the time difference between any two intervals on the lightcurve. As described in \cite{koz16}, the time lag is defined as $\Delta$T = t${_i}$-t${_j}$ where i and j are any two different epochs on the lightcurve. We correct for the observed frame variability to rest frame by dividing $\Delta$T by (1+z). We use the spectroscopic redshifts for our sources when known and otherwise use the cluster redshift for all other sources. The Structure Function (SF) is given in the equation below. We have adopted the version of the SF used by \cite{vb04}, but adjusted for the incomplete noise term by subtracting 2$\sigma$ as discussed in \cite{koz16}.

\[SF(\Delta T) = \sqrt{\frac{\pi}{2} \langle|m_i-m_j|\rangle^2 - 2\langle\sigma^2\rangle}\]

\begin{figure}
\centering
\includegraphics[width=\textwidth]{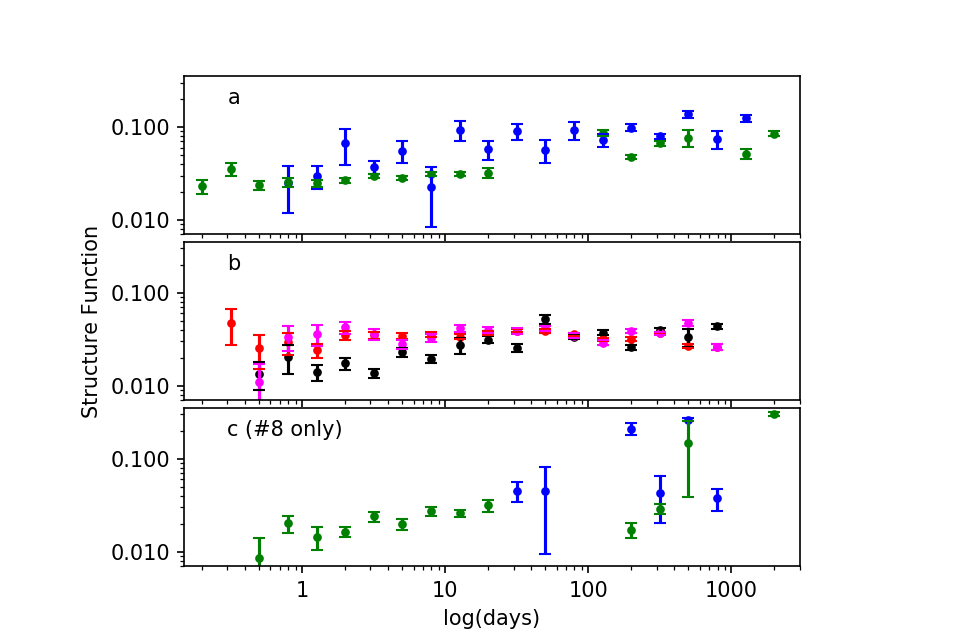}
\caption{The Structure Function vs. time lag for all optical variables (a) and all IR variables (b). V-band variables are shown in blue, I-band variables in green. In the IR, F105W variables are shown in black, F125W variables are red, and F160W variables are in magenta. c) The Structure Function for source \#8 in the V-band (blue) and I-band (green).}
\label{fig:structure_functions}
\end{figure}

Figure\,~\ref{fig:structure_functions} is the SF for the optical (a) and IR (b) lightcurves using all sources detected as significant variables in each band. We find that the overall amplitude of the SF is greater at shorter wavelengths with the most variability found in the V-band and decreasing towards the IR bands. This anti-correlation with wavelength has been observed in previous studies of AGN (\citealt{unai14, vb04, zuo12}). Generally, the SF for AGN is found to rise as a power-law towards longer wavelengths, reaching a peak variability amplitude at some characteristic time-scale. \cite{koz16} analyzed $\sim$9000 SDSS r-band quasar lightcurves and found that the SF slope was around 0.5, consistent with that for the Damped Random Walk (DRW) model \citep{kelly14}. The characteristic timescale where the SF flattens was found to be at $\sim$1 year. While our SF extends beyond one year, we do not find any evidence of flattening at these timescales. However, we note that the noise in the SF at larger timescales may prevent us from accurately determining the characteristic timescale for our sources. 

For simplicity, we fit a single power-law to the SF and measure the following slopes; 0.175 $\pm$0.020 (F606W), 0.156 $\pm$0.007 (F814W), and 0.148 $\pm$0.013 (F105W). The F125W and F160W SFs are consistent with being flat. We compare our slopes with the single power-law fit to the ensemble SF of Seyfert-like AGNs \citep{unai14} in the SDSS. They find slopes between 0.22 and 0.24 in the three SDSS optical filters. The SF slopes for our sources are shallower than these and much shallower than the SF slope of SDSS quasars presented in \cite{koz16}. The reason for this is likely due to the presence of the host galaxy in our photometry. The galaxies make a non-negligible contribution to the optical and IR bands by adding to the total flux but not to the fractional flux variations. This decreases the amplitude of the observed variability and diminishes the power of the SF. In fact, we would expect to see greater effects for longer wavelengths, which would explain the lower overall flux amplitude with increasing wavelength as well as the weakened SF slope with increasing wavelength.

To further test this assertion, we compute the SF for source \#8. This object is the only variable in our survey with significant variability at all 5 wavelengths. We computed the SF for this source using the V and I band lightcurves (Figure\,~\ref{fig:structure_functions}c). The lightcurves are very poorly sampled in the IR bands (6 or fewer epochs) which prevented an accurate SF from being constructed at these wavelengths. We find that the flattening at large time intervals is again difficult to determine with our data, so we fit a single power-law through the SF and find slopes of 0.466 $\pm$ 0.078 in V and 0.416 $\pm$ 0.013 in I. These slopes are much closer to that found for quasars and the DRW model slope of 0.5. We also find that the variability amplitudes at the largest time lags for this source (1 to 3 years) are close to the peak amplitude for SDSS quasars found by \cite{koz16} of $\sim$0.2 mag. These results are consistent with the interpretation that \#8 is dominated by the AGN component. Therefore, the contributions of the host galaxy are negligible and the SF values are much closer to those found for large samples of quasars. 

\section{Conclusions} \label{sec:highlight}

The cluster MACSJ1149 is an ideal candidate for variability investigations due to the vast amount of {\sl HST} images obtained in this field over long and short time scales. Our variability analysis detected 49 candidate AGN and 4 previously undetected SNe. Out of the $\sim$2000 galaxies included in our variability analysis, and taking into account spurious sources, the fraction of variability selected AGN detected in any wavelength is about 2\%, which is consistent with other {\sl HST} variability studies. 

We find that 55\% of the variables are in the cluster, with the majority of non-cluster sources (39\% of all variables) beyond the cluster redshift of z=0.55. Variable AGN candidates make up a larger fraction of cluster galaxies (6.5\%) compared to those found among foreground and background galaxies (1.3\%). Of the variables detected only in the NIR bands, the majority (70\%) are in the cluster and have morphologies consistent with being elliptical galaxies. NIR variability may provide an important tool for identifying AGN in these redder galaxies. Among those galaxies not in the cluster, very few have elliptical morphologies or elliptical colors and almost half have point-like morphologies. About 20\% of the sources have BVI colors consistent with typical Type-1 AGN, indicating that the majority of our variables are dominated by the host galaxy. We also find that galaxies hosting a variable nucleus tend to be among the more luminous galaxies at all redshifts. 

The fraction of cluster galaxies hosting a variable nucleus is greater towards the cluster center, about 10\%, and decreases to about 3\% at one-third of the virial radius of the cluster. This trend is consistent with previous studies of X-ray and mid-IR selected AGNs, which are more common in the centers of clusters. These results support moderate ram pressure stripping and/or galaxy harassment as effective triggers for AGN accretion.

We compute the Structure Function for the variable AGN candidates in our survey. We find an anti-correlation between the overall variability amplitude and wavelength. The Structure Function slopes are much shallower than those found for SDSS quasars, which may be due to the host galaxy light decreasing the amplitude of the observed variability for our sources. The Structure Function for the most variable source in our sample has a much steeper slope, approaching the slope predicted by the Damped Random Walk model.

We obtained follow-up spectroscopy for six of our variables using the GTC and OSIRIS MOS. Two of our sources revealed broad emission lines from which we measured the black hole masses and accretion rates. We have also discovered a previously unknown lensed quasar in the form of an Einstein cross as well as 4 new supernovae, doubling the previous SN rate in this field.

\section*{Acknowledgement}
This research was based on observations made with the NASA/ESA Hubble Space Telescope, and from the data archive at the Space Telescope Science Institute. STScI is operated by the Association of Universities for Research in Astronomy, Inc. under NASA contract NAS 5-26555. This research also made use of observations made with the Gran Telescopio Canarias (GTC), installed at the Spanish Observatorio del Roque de los Muchachos of the Instituto de Astrofísica de Canarias, in the island of La Palma.

\bibliography{refereces}

\end{document}

%% file: tables/filter_data_table.tex
$F435W$ &  11   & 0.30 days - 4.21 years  & 4062.18 \\
$F606W$ &  25   & 0.73 days - 6.48 years  & 2012.12 \\
$F814W$ &  26   & 0.07 days - 12.1 years  & 4457.38 \\
$F105W$ &  27   & 0.52 days - 5.30 years  & 3261.00 \\
$F125W$ &  71   & 0.27 days - 6.09 years  & 1541.00 \\
$F160W$ &  72   & 0.59 days - 6.01 years  & 1909.00 \\

%% file: tables/variable_table.tex
1  & 177.367279 & 22.397551 & 0.454 & -     & 4.99  & -     & -     & - & 0\\
2  & 177.370468 & 22.391569 & 0.503 & -     & 3.79  & -     & -     & - & 0\\
3  & 177.414734 & 22.419691 & 2.301 & 8.02  & 4.99  & -     & -     & - & 2\\
4  & 177.400497 & 22.398207 & 0.541 & -     & 3.09  & -     & -     & - & 0\\
5  & 177.401810 & 22.398804 & 0.543 & -     & 3.77  & -     & -     & - & 0\\
6  & 177.394501 & 22.389168 & 0.962 & 3.89  & 3.97  & 4.45  & -     & - & 0\\
7  & 177.415222 & 22.406197 & 1.683 & -     & 4.23  & -     & -     & - & 4\\
8  & 177.391266 & 22.374170 & 0.555 & 12.16 & 36.18 & 50.29 & 40.97 & 10.88 & 3\\
9  & 177.402710 & 22.384066 & 4.406 & -     & 3.59  & -     & -     & - & 2\\
10 & 177.424301 & 22.386259 & 3.448 & 17.01 & 5.69  & -     & -     & - & 3\\
11 & 177.416336 & 22.392332 & 0.542 & -     & 3.95  & -     & -     & - & 0\\
12 & 177.417694 & 22.371298 & 4.183 & -     & 3.98  & -     & -     & - & 2\\
13 & 177.411682 & 22.429659 & 1.227 & 89.82 & -     & -     & -     & - & 3\\
14 & 177.378479 & 22.390856 & 0.371 & 7.69  & -     & 8.64  & -     & - & 1\\
15 & 177.418198 & 22.425699 & 0.639 & 4.93  & -     & -     & -     & - & 2\\
16 & 177.387970 & 22.384907 & 0.107 & 4.89  & -     & -     & -     & - & 2\\
17 & 177.431381 & 22.404469 & 1.371 & 6.75  & -     & -     & -     & - & 3\\
18 & 177.404480 & 22.382647 & 2.693 & 3.13  & -     & -     & -     & - & 4\\
19 & 177.387833 & 22.376003 & 0.525 & 3.14  & -     & -     & -     & - & 2\\
20 & 177.433533 & 22.387032 & 0.545 & 5.71  & -     & -     & -     & - & 1\\
21 & 177.422546 & 22.385674 & 1.976 & 4.36  & -     & -     & -     & - & 4\\
22 & 177.431015 & 22.381664 & 2.367 & 8.31  & -     & -     & -     & - & 4\\
23 & 177.396683 & 22.413326 & 0.559 & -     & -     & 5.41  & -     & 5.93 & 0\\
24 & 177.388123 & 22.411812 & 0.541 & -     & -     & 7.06  & -     & - & 0\\
25 & 177.401093 & 22.404703 & 1.086 & -     & -     & 4.7   & -     & - & 1\\
26 & 177.390244 & 22.403900 & 0.543 & -     & -     & 7.67  & -     & - & 0\\
27 & 177.393555 & 22.401325 & 0.723 & -     & -     & 6.11  & -     & - & 0\\
28 & 177.391479 & 22.401068 & 0.538 & -     & -     & 5.26  & -     & - & 0\\
29 & 177.384811 & 22.400492 & 0.547 & -     & -     & 8.68  & -     & - & 0\\
30 & 177.412094 & 22.397606 & 0.496 & -     & -     & 7.78  & -     & - & 0\\
31 & 177.384079 & 22.384739 & 0.539 & -     & -     & 3.47  & -     & 3.52 & 0\\
32 & 177.388580 & 22.386656 & 0.230 & -     & -     & 3.03  & -     & - & 0/4\\
33 & 177.396881 & 22.378502 & 0.485 & -     & -     & 3.3   & -     & - & 0\\
34 & 177.395889 & 22.418217 & 0.556 & -     & -     & -     & 3.23  & - & 0\\
35 & 177.379593 & 22.411879 & 3.078 & -     & -     & -     & 14.64 & 8.62 & 2\\
36 & 177.398361 & 22.410786 & 0.563 & -     & -     & -     & 4.81  & - & 0\\
37 & 177.394150 & 22.408436 & 0.540 & -     & -     & -     & 5.23  & - & 0\\
38 & 177.373840 & 22.404655 & 0.544 & -     & -     & -     & 4.86  & 10.01 & 0\\
39 & 177.403152 & 22.404385 & 0.534 & -     & -     & -     & 7.35  & - & 0\\
40 & 177.393906 & 22.402308 & 0.538 & -     & -     & -     & 3.15  & - & 0\\
41 & 177.398956 & 22.401814 & 0.542 & -     & -     & -     & 5.83  & 4.49 & 0\\
42 & 177.401123 & 22.397884 & 0.541 & -     & -     & -     & 3.42  & - & 0\\
43 & 177.424118 & 22.394613 & 0.748 & -     & -     & -     & 5.01  & - & 0\\
44 & 177.403778 & 22.388393 & 0.533 & -     & -     & -     & 4.63  & - & 0\\
45 & 177.389328 & 22.374210 & 1.646 & -     & -     & -     & 3.92  & - & 4\\
46 & 177.379440 & 22.408911 & 0.556 & -     & -     & -     & -     & 6.06 & 0\\
47 & 177.420120 & 22.398260 & 0.962 & -     & -     & -     & -     & 11.26 & 2\\
48 & 177.403748 & 22.391943 & 0.544 & -     & -     & -     & -     & 3.36 & 0\\
49 & 177.380692 & 22.381935 & 0.636 & -     & -     & -     & -     & 3.5  & 0\\

%% file: tables/sn_table.tex


$Dellanova$ & 177.431818 & 22.407301 & F814W       & 4/2015 & 26.24       & 25 \\
$Saranova$  & 177.416129 & 22.382253 & F125W;F160W & 2/2016 & 24.84;24.61 & 17 \\
$C1$        & 177.380125 & 22.381277 & F814W       & 4/2004 & 24.95       & 24 \\
$C2$        & 177.398944 & 22.371776 & F814W       & 5/2006 & 25.18       & 24 \\

%% file: AGN_Paper_revisions_Jan2020.bbl
\begin{thebibliography}{}
\expandafter\ifx\csname natexlab\endcsname\relax\def\natexlab#1{#1}\fi
\providecommand{\url}[1]{\href{#1}{#1}}

\bibitem[{{Abell} {et~al.}(1989){Abell}, {Corwin}, \& {Olowin}}]{abell89}
{Abell}, G.~O., {Corwin}, Harold~G., J., \& {Olowin}, R.~P. 1989, \apjs, 70, 1

\bibitem[{{Aguado} {et~al.}(2019){Aguado}, {Ahumada}, {Almeida}, {Anderson},
  {Andrews}, {Anguiano}, {Aquino Ort{\'\i}z}, {Arag{\'o}n-Salamanca},
  {Argudo-Fern{\'a}ndez}, {Aubert}, {Avila-Reese}, {Badenes}, {Barboza
  Rembold}, {Barger}, {Barrera-Ballesteros}, {Bates}, {Bautista}, {Beaton},
  {Beers}, {Belfiore}, {Bernardi}, {Bershady}, {Beutler}, {Bird}, {Bizyaev},
  {Blanc}, {Blanton}, {Blomqvist}, {Bolton}, {Boquien}, {Borissova}, {Bovy},
  {Brand t}, {Brinkmann}, {Brownstein}, {Bundy}, {Burgasser}, {Byler}, {Cano
  Diaz}, {Cappellari}, {Carrera}, {Cervantes Sodi}, {Chen}, {Cherinka}, {Choi},
  {Chung}, {Coffey}, {Comerford}, {Comparat}, {Covey}, {da Silva Ilha}, {da
  Costa}, {Dai}, {Damke}, {Darling}, {Davies}, {Dawson}, {de Sainte Agathe},
  {Deconto Machado}, {Del Moro}, {De Lee}, {Diamond-Stanic}, {Dom{\'\i}nguez
  S{\'a}nchez}, {Donor}, {Drory}, {du Mas des Bourboux}, {Duckworth}, {Dwelly},
  {Ebelke}, {Emsellem}, {Escoffier}, {Fern{\'a}ndez-Trincado}, {Feuillet},
  {Fischer}, {Fleming}, {Fraser-McKelvie}, {Freischlad}, {Frinchaboy}, {Fu},
  {Galbany}, {Garcia-Dias}, {Garc{\'\i}a-Hern{\'a}ndez}, {Garma Oehmichen},
  {Geimba Maia}, {Gil-Mar{\'\i}n}, {Grabowski}, {Gu}, {Guo}, {Ha},
  {Harrington}, {Hasselquist}, {Hayes}, {Hearty}, {Hernandez Toledo}, {Hicks},
  {Hogg}, {Holley-Bockelmann}, {Holtzman}, {Hsieh}, {Hunt}, {Hwang},
  {Ibarra-Medel}, {Jimenez Angel}, {Johnson}, {Jones}, {J{\"o}nsson},
  {Kinemuchi}, {Kollmeier}, {Krawczyk}, {Kreckel}, {Kruk}, {Lacerna}, {Lan},
  {Lane}, {Law}, {Lee}, {Li}, {Lian}, {Lin}, {Lin}, {Lintott}, {Long},
  {Longa-Pe{\~n}a}, {Mackereth}, {de la Macorra}, {Majewski}, {Malanushenko},
  {Manchado}, {Maraston}, {Mariappan}, {Marinelli}, {Marques-Chaves},
  {Masseron}, {Masters}, {McDermid}, {Medina Pe{\~n}a}, {Meneses-Goytia},
  {Merloni}, {Merrifield}, {Meszaros}, {Minniti}, {Minsley}, {Muna}, {Myers},
  {Nair}, {Correa do Nascimento}, {Newman}, {Nitschelm}, {Olmstead}, {Oravetz},
  {Oravetz}, {Ortega Minakata}, {Pace}, {Padilla}, {Palicio}, {Pan}, {Pan},
  {Parikh}, {Parker}, {Peirani}, {Penny}, {Percival}, {Perez-Fournon},
  {Peterken}, {Pinsonneault}, {Prakash}, {Raddick}, {Raichoor}, {Riffel},
  {Riffel}, {Rix}, {Robin}, {Roman-Lopes}, {Rose}, {Ross}, {Rossi}, {Rowlands},
  {Rubin}, {S{\'a}nchez}, {S{\'a}nchez-Gallego}, {Sayres}, {Schaefer},
  {Schiavon}, {Schimoia}, {Schlafly}, {Schlegel}, {Schneider}, {Schultheis},
  {Seo}, {Shamsi}, {Shao}, {Shen}, {Shetty}, {Simonian}, {Smethurst}, {Sobeck},
  {Souter}, {Spindler}, {Stark}, {Stassun}, {Steinmetz}, {Storchi-Bergmann},
  {Stringfellow}, {Su{\'a}rez}, {Sun}, {Taghizadeh-Popp}, {Talbot}, {Tayar},
  {Thakar}, {Thomas}, {Tissera}, {Tojeiro}, {Troup}, {Unda-Sanzana},
  {Valenzuela}, {Vargas-Maga{\~n}a}, {V{\'a}zquez-Mata}, {Wake}, {Weaver},
  {Weijmans}, {Westfall}, {Wild}, {Wilson}, {Woods}, {Yan}, {Yang}, {Zamora},
  {Zasowski}, {Zhang}, {Zheng}, {Zheng}, {Zhu}, {Zinn}, \& {Zou}}]{aguado19}
{Aguado}, D.~S., {Ahumada}, R., {Almeida}, A., {et~al.} 2019, \apjs, 240, 23

\bibitem[{{Alard} \& {Lupton}(1998)}]{alard98}
{Alard}, C., \& {Lupton}, R.~H. 1998, \apj, 503, 325

\bibitem[{{Alexander} {et~al.}(2003){Alexander}, {Bauer}, {Brandt},
  {Schneider}, {Hornschemeier}, {Vignali}, {Barger}, {Broos}, {Cowie},
  {Garmire}, {Townsley}, {Bautz}, {Chartas}, \& {Sargent}}]{alexander03}
{Alexander}, D.~M., {Bauer}, F.~E., {Brandt}, W.~N., {et~al.} 2003, AJ, 126,
  539

\bibitem[{{Bahk} {et~al.}(2019){Bahk}, {Woo}, \& {Park}}]{bahk19}
{Bahk}, H., {Woo}, J.-H., \& {Park}, D. 2019, \apj, 875, 50

\bibitem[{{Baldwin} {et~al.}(1981){Baldwin}, {Phillips}, \&
  {Terlevich}}]{baldwin81}
{Baldwin}, J.~A., {Phillips}, M.~M., \& {Terlevich}, R. 1981, PASP, 93, 5

\bibitem[{{Bershady} {et~al.}(1998){Bershady}, {Trevese}, \&
  {Kron}}]{bershady98}
{Bershady}, M.~A., {Trevese}, D., \& {Kron}, R.~G. 1998, \apj, 496, 103

\bibitem[{{Bertin} \& {Arnouts}(1996)}]{bertin96}
{Bertin}, E., \& {Arnouts}, S. 1996, \aaps, 117, 393

\bibitem[{{Bettoni} {et~al.}(2019){Bettoni}, {Falomo}, {Scarpa}, {Negrello},
  {Omizzolo}, {Corradi}, {Reverte}, \& {Vulcani}}]{bettoni19}
{Bettoni}, D., {Falomo}, R., {Scarpa}, R., {et~al.} 2019, \apjl, 873, L14

\bibitem[{{Coe} {et~al.}(2006){Coe}, {Ben{\'\i}tez}, {S{\'a}nchez}, {Jee},
  {Bouwens}, \& {Ford}}]{coe06}
{Coe}, D., {Ben{\'\i}tez}, N., {S{\'a}nchez}, S.~F., {et~al.} 2006, \aj, 132,
  926

\bibitem[{{Ebeling} {et~al.}(2001){Ebeling}, {Edge}, \& {Henry}}]{ebeling01}
{Ebeling}, H., {Edge}, A.~C., \& {Henry}, J.~P. 2001, \apj, 553, 668

\bibitem[{{Ebeling} {et~al.}(2014){Ebeling}, {Ma}, \& {Barrett}}]{ebeling14}
{Ebeling}, H., {Ma}, C.-J., \& {Barrett}, E. 2014, \apjs, 211, 21

\bibitem[{{Ehlert} {et~al.}(2013){Ehlert}, {Werner}, {Simionescu}, {Allen},
  {Kenney}, {Million}, \& {Finoguenov}}]{elhert13}
{Ehlert}, S., {Werner}, N., {Simionescu}, A., {et~al.} 2013, \mnras, 430, 2401

\bibitem[{{Fabbiano} {et~al.}(1992){Fabbiano}, {Kim}, \&
  {Trinchieri}}]{fabbiano92}
{Fabbiano}, G., {Kim}, D.-W., \& {Trinchieri}, G. 1992, ApJS, 80, 531

\bibitem[{{Fassbender} {et~al.}(2012){Fassbender}, {{\v{S}}uhada}, \&
  {Nastasi}}]{fassbender12}
{Fassbender}, R., {{\v{S}}uhada}, R., \& {Nastasi}, A. 2012, Advances in
  Astronomy, 2012, 138380

\bibitem[{{Ferrarese} \& {Merritt}(2000)}]{ferrarese00}
{Ferrarese}, L., \& {Merritt}, D. 2000, \apjl, 539, L9

\bibitem[{{Galametz} {et~al.}(2009){Galametz}, {Stern}, {Eisenhardt},
  {Brodwin}, {Brown}, {Dey}, {Gonzalez}, {Jannuzi}, {Moustakas}, \&
  {Stanford}}]{galametz09}
{Galametz}, A., {Stern}, D., {Eisenhardt}, P. R.~M., {et~al.} 2009, \apj, 694,
  1309

\bibitem[{{Gallastegui-Aizpun} \& {Sarajedini}(2014)}]{unai14}
{Gallastegui-Aizpun}, U., \& {Sarajedini}, V.~L. 2014, \mnras, 444, 3078

\bibitem[{{Gebhardt} {et~al.}(2000){Gebhardt}, {Bender}, {Bower}, {Dressler},
  {Faber}, {Filippenko}, {Green}, {Grillmair}, {Ho}, {Kormendy}, {Lauer},
  {Magorrian}, {Pinkney}, {Richstone}, \& {Tremaine}}]{gebhardt00}
{Gebhardt}, K., {Bender}, R., {Bower}, G., {et~al.} 2000, \apjl, 539, L13

\bibitem[{{Gonzaga} \& {et al.}(2012)}]{gonzaga12}
{Gonzaga}, S., \& {et al.} 2012, {The DrizzlePac Handbook}

\bibitem[{{Grillo}(2015)}]{grillo15}
{Grillo}, C. 2015, in IAU General Assembly, Vol.~29, 2256714

\bibitem[{{Hart} {et~al.}(2011){Hart}, {Stocke}, {Evrard}, {Ellingson}, \&
  {Barkhouse}}]{hart11}
{Hart}, Q.~N., {Stocke}, J.~T., {Evrard}, A.~E., {Ellingson}, E.~E., \&
  {Barkhouse}, W.~A. 2011, \apj, 740, 59

\bibitem[{{Hickox} {et~al.}(2009){Hickox}, {Jones}, {Forman}, {Murray},
  {Kochanek}, {Eisenstein}, {Jannuzi}, {Dey}, {Brown}, {Stern}, {Eisenhardt},
  {Gorjian}, {Brodwin}, {Narayan}, {Cool}, {Kenter}, {Caldwell}, \&
  {Anderson}}]{hickox09}
{Hickox}, R.~C., {Jones}, C., {Forman}, W.~R., {et~al.} 2009, ApJ, 696, 891

\bibitem[{{Hillebrandt} {et~al.}(2000){Hillebrandt}, {Reinecke}, \&
  {Niemeyer}}]{hill00}
{Hillebrandt}, W., {Reinecke}, M., \& {Niemeyer}, J.~C. 2000, arXiv e-prints,
  astro

\bibitem[{{Hook} {et~al.}(1994){Hook}, {McMahon}, {Boyle}, \& {Irwin}}]{hook94}
{Hook}, I.~M., {McMahon}, R.~G., {Boyle}, B.~J., \& {Irwin}, M.~J. 1994, MNRAS,
  268, 305

\bibitem[{{Hopkins} \& {Hernquist}(2006)}]{hopkins06}
{Hopkins}, P.~F., \& {Hernquist}, L. 2006, \apjs, 166, 1

\bibitem[{{Kelly} {et~al.}(2009){Kelly}, {Bechtold}, \&
  {Siemiginowska}}]{kelly09}
{Kelly}, B.~C., {Bechtold}, J., \& {Siemiginowska}, A. 2009, \apj, 698, 895

\bibitem[{{Kelly} {et~al.}(2014){Kelly}, {Becker}, {Sobolewska},
  {Siemiginowska}, \& {Uttley}}]{kelly14}
{Kelly}, B.~C., {Becker}, A.~C., {Sobolewska}, M., {Siemiginowska}, A., \&
  {Uttley}, P. 2014, \apj, 788, 33

\bibitem[{{Kelly} {et~al.}(2011){Kelly}, {Sobolewska}, \&
  {Siemiginowska}}]{kelly11}
{Kelly}, B.~C., {Sobolewska}, M., \& {Siemiginowska}, A. 2011, \apj, 730, 52

\bibitem[{{Kelly} {et~al.}(2017){Kelly}, {Diego}, {Nonino}, {Zitrin}, {Jauzac},
  \& {Filippenko}}]{kelly17}
{Kelly}, P.~L., {Diego}, J.~M., {Nonino}, M., {et~al.} 2017, The Astronomer's
  Telegram, 10005, 1

\bibitem[{{Kelly} {et~al.}(2015){Kelly}, {Rodney}, {Treu}, {Foley}, {Brammer},
  {Schmidt}, {Zitrin}, {Sonnenfeld}, {Strolger}, {Graur}, {Filippenko}, {Jha},
  {Riess}, {Bradac}, {Weiner}, {Scolnic}, {Malkan}, {von der Linden}, {Trenti},
  {Hjorth}, {Gavazzi}, {Fontana}, {Merten}, {McCully}, {Jones}, {Postman},
  {Dressler}, {Patel}, {Cenko}, {Graham}, \& {Tucker}}]{kelly15}
{Kelly}, P.~L., {Rodney}, S.~A., {Treu}, T., {et~al.} 2015, Science, 347, 1123

\bibitem[{{Kelly} {et~al.}(2016){Kelly}, {Rodney}, {Treu}, {Strolger}, {Foley},
  {Jha}, {Selsing}, {Brammer}, {Brada{\v{c}}}, {Cenko}, {Graur}, {Filippenko},
  {Hjorth}, {McCully}, {Molino}, {Nonino}, {Riess}, {Schmidt}, {Tucker}, {von
  der Linden}, {Weiner}, \& {Zitrin}}]{kelly16}
---. 2016, \apjl, 819, L8

\bibitem[{{Kelly} {et~al.}(2018){Kelly}, {Diego}, {Rodney}, {Kaiser},
  {Broadhurst}, {Zitrin}, {Treu}, {P{\'e}rez-Gonz{\'a}lez}, {Morishita},
  {Jauzac}, {Selsing}, {Oguri}, {Pueyo}, {Ross}, {Filippenko}, {Smith},
  {Hjorth}, {Cenko}, {Wang}, {Howell}, {Richard}, {Frye}, {Jha}, {Foley},
  {Norman}, {Bradac}, {Zheng}, {Brammer}, {Benito}, {Cava}, {Christensen}, {de
  Mink}, {Graur}, {Grillo}, {Kawamata}, {Kneib}, {Matheson}, {McCully},
  {Nonino}, {P{\'e}rez-Fournon}, {Riess}, {Rosati}, {Schmidt}, {Sharon}, \&
  {Weiner}}]{kelly18}
{Kelly}, P.~L., {Diego}, J.~M., {Rodney}, S., {et~al.} 2018, Nature Astronomy,
  2, 334

\bibitem[{{Klesman} \& {Sarajedini}(2014)}]{klesman14}
{Klesman}, A.~J., \& {Sarajedini}, V.~L. 2014, \mnras, 442, 314

\bibitem[{{Koo} {et~al.}(1986){Koo}, {Kron}, \& {Cudworth}}]{kkc86}
{Koo}, D.~C., {Kron}, R.~G., \& {Cudworth}, K.~M. 1986, PASP, 98, 285

\bibitem[{{Kormendy} \& {Richstone}(1995)}]{kormendy95}
{Kormendy}, J., \& {Richstone}, D. 1995, \araa, 33, 581

\bibitem[{{Koz{\l}owski}(2016)}]{koz16}
{Koz{\l}owski}, S. 2016, \apj, 826, 118

\bibitem[{{Lacy} {et~al.}(2004){Lacy}, {Storrie-Lombardi}, {Sajina},
  {Appleton}, {Armus}, {Chapman}, {Choi}, {Fadda}, {Fang}, {Frayer},
  {Heinrichsen}, {Helou}, {Im}, {Marleau}, {Masci}, {Shupe}, {Soifer},
  {Surace}, {Teplitz}, {Wilson}, \& {Yan}}]{lacy04}
{Lacy}, M., {Storrie-Lombardi}, L.~J., {Sajina}, A., {et~al.} 2004, ApJS, 154,
  166

\bibitem[{{Lira} {et~al.}(2015){Lira}, {Ar{\'e}valo}, {Uttley}, {McHardy}, \&
  {Videla}}]{lira15}
{Lira}, P., {Ar{\'e}valo}, P., {Uttley}, P., {McHardy}, I.~M.~M., \& {Videla},
  L. 2015, \mnras, 454, 368

\bibitem[{{Lotz} {et~al.}(2017){Lotz}, {Koekemoer}, {Coe}, {Grogin}, {Capak},
  {Mack}, {Anderson}, {Avila}, {Barker}, {Borncamp}, {Brammer}, {Durbin},
  {Gunning}, {Hilbert}, {Jenkner}, {Khandrika}, {Levay}, {Lucas}, {MacKenty},
  {Ogaz}, {Porterfield}, {Reid}, {Robberto}, {Royle}, {Smith},
  {Storrie-Lombardi}, {Sunnquist}, {Surace}, {Taylor}, {Williams}, {Bullock},
  {Dickinson}, {Finkelstein}, {Natarajan}, {Richard}, {Robertson}, {Tumlinson},
  {Zitrin}, {Flanagan}, {Sembach}, {Soifer}, \& {Mountain}}]{lotz17}
{Lotz}, J.~M., {Koekemoer}, A., {Coe}, D., {et~al.} 2017, \apj, 837, 97

\bibitem[{{MacLeod} {et~al.}(2010){MacLeod}, {Ivezi{\'c}}, {Kochanek},
  {Koz{\l}owski}, {Kelly}, {Bullock}, {Kimball}, {Sesar}, {Westman}, {Brooks},
  {Gibson}, {Becker}, \& {de Vries}}]{macleod10}
{MacLeod}, C.~L., {Ivezi{\'c}}, {\v Z}., {Kochanek}, C.~S., {et~al.} 2010, ApJ,
  721, 1014

\bibitem[{{Markarian}(1967)}]{mark67}
{Markarian}, B.~E. 1967, Astrofizika, 3, 55

\bibitem[{{Marshall} {et~al.}(2018){Marshall}, {Shabala}, {Krause}, {Pimbblet},
  {Croton}, \& {Owers}}]{marshall18}
{Marshall}, M.~A., {Shabala}, S.~S., {Krause}, M. G.~H., {et~al.} 2018, \mnras,
  474, 3615

\bibitem[{{Mo} {et~al.}(2018){Mo}, {Gonzalez}, {Stern}, {Brodwin}, {Decker},
  {Eisenhardt}, {Moravec}, {Stanford}, \& {Wylezalek}}]{mo18}
{Mo}, W., {Gonzalez}, A., {Stern}, D., {et~al.} 2018, \apj, 869, 131

\bibitem[{{Molino} {et~al.}(2017){Molino}, {Ben{\'\i}tez}, {Ascaso}, {Coe},
  {Postman}, {Jouvel}, {Host}, {Lahav}, {Seitz}, {Medezinski}, {Rosati},
  {Schoenell}, {Koekemoer}, {Jimenez-Teja}, {Broadhurst}, {Melchior},
  {Balestra}, {Bartelmann}, {Bouwens}, {Bradley}, {Czakon}, {Donahue}, {Ford},
  {Graur}, {Graves}, {Grillo}, {Infante}, {Jha}, {Kelson}, {Lazkoz}, {Lemze},
  {Maoz}, {Mercurio}, {Meneghetti}, {Merten}, {Moustakas}, {Nonino}, {Orgaz},
  {Riess}, {Rodney}, {Sayers}, {Umetsu}, {Zheng}, \& {Zitrin}}]{molino17}
{Molino}, A., {Ben{\'\i}tez}, N., {Ascaso}, B., {et~al.} 2017, \mnras, 470, 95

\bibitem[{{Pereyra} {et~al.}(2006){Pereyra}, {Vanden Berk}, {Turnshek},
  {Hillier}, {Wilhite}, {Kron}, {Schneider}, \& {Brinkmann}}]{pereyra06}
{Pereyra}, N.~A., {Vanden Berk}, D.~E., {Turnshek}, D.~A., {et~al.} 2006, ApJ,
  642, 87

\bibitem[{{Peterson} {et~al.}(1998){Peterson}, {Wanders}, {Bertram}, {Hunley},
  {Pogge}, \& {Wagner}}]{peterson98}
{Peterson}, B.~M., {Wanders}, I., {Bertram}, R., {et~al.} 1998, \apj, 501, 82

\bibitem[{{Postman} {et~al.}(2012){Postman}, {Coe}, {Ben{\'\i}tez}, {Bradley},
  {Broadhurst}, {Donahue}, {Ford}, {Graur}, {Graves}, {Jouvel}, {Koekemoer},
  {Lemze}, {Medezinski}, {Molino}, {Moustakas}, {Ogaz}, {Riess}, {Rodney},
  {Rosati}, {Umetsu}, {Zheng}, {Zitrin}, {Bartelmann}, {Bouwens}, {Czakon},
  {Golwala}, {Host}, {Infante}, {Jha}, {Jimenez-Teja}, {Kelson}, {Lahav},
  {Lazkoz}, {Maoz}, {McCully}, {Melchior}, {Meneghetti}, {Merten}, {Moustakas},
  {Nonino}, {Patel}, {Reg{\"o}s}, {Sayers}, {Seitz}, \& {Van der
  Wel}}]{postman12}
{Postman}, M., {Coe}, D., {Ben{\'\i}tez}, N., {et~al.} 2012, \apjs, 199, 25

\bibitem[{{Pouliasis} {et~al.}(2019){Pouliasis}, {Georgantopoulos}, {Bonanos},
  {Yang}, {Sokolovsky}, {Hatzidimitriou}, {Mountrichas}, {Gavras},
  {Charmandaris}, {Bellas-Velidis}, {Spetsieri}, \& {Tsinganos}}]{pouliasis19}
{Pouliasis}, E., {Georgantopoulos}, I., {Bonanos}, A.~Z., {et~al.} 2019,
  \mnras, 487, 4285

\bibitem[{{Rawle} {et~al.}(2016){Rawle}, {Altieri}, {Egami},
  {P{\'e}rez-Gonz{\'a}lez}, {Boone}, {Clement}, {Ivison}, {Richard},
  {Rujopakarn}, {Valtchanov}, {Walth}, {Weiner}, {Blain}, {Dessauges-Zavadsky},
  {Kneib}, {Lutz}, {Rodighiero}, {Schaerer}, \& {Smail}}]{rawle16}
{Rawle}, T.~D., {Altieri}, B., {Egami}, E., {et~al.} 2016, \mnras, 459, 1626

\bibitem[{{Richards} {et~al.}(2002){Richards}, {Fan}, {Newberg}, {Strauss},
  {Vanden Berk}, {Schneider}, {Yanny}, {Boucher}, {Burles}, {Frieman}, {Gunn},
  {Hall}, {Ivezi{\'c}}, {Kent}, {Loveday}, {Lupton}, {Rockosi}, {Schlegel},
  {Stoughton}, {SubbaRao}, \& {York}}]{richards02}
{Richards}, G.~T., {Fan}, X., {Newberg}, H.~J., {et~al.} 2002, AJ, 123, 2945

\bibitem[{{Ruan} {et~al.}(2014){Ruan}, {Anderson}, {Dexter}, \&
  {Agol}}]{ruan14}
{Ruan}, J.~J., {Anderson}, S.~F., {Dexter}, J., \& {Agol}, E. 2014, \apj, 783,
  105

\bibitem[{{Sarajedini} {et~al.}(2011){Sarajedini}, {Koo}, {Klesman}, {Laird},
  {Perez Gonzalez}, \& {Mozena}}]{saraj11}
{Sarajedini}, V.~L., {Koo}, D.~C., {Klesman}, A.~J., {et~al.} 2011, \apj, 731,
  97

\bibitem[{{Schmidt} {et~al.}(2010){Schmidt}, {Marshall}, {Rix}, {Jester},
  {Hennawi}, \& {Dobler}}]{Schmidt10}
{Schmidt}, K.~B., {Marshall}, P.~J., {Rix}, H.-W., {et~al.} 2010, \apj, 714,
  1194

\bibitem[{{Schmidt} \& {Green}(1983)}]{sg83}
{Schmidt}, M., \& {Green}, R.~F. 1983, ApJ, 269, 352

\bibitem[{{Shipley} {et~al.}(2018){Shipley}, {Lange-Vagle}, {Marchesini},
  {Brammer}, {Ferrarese}, {Stefanon}, {Kado-Fong}, {Whitaker}, {Oesch},
  {Feinstein}, {Labb{\'e}}, {Lundgren}, {Martis}, {Muzzin}, {Nedkova},
  {Skelton}, \& {van der Wel}}]{shipley18}
{Shipley}, H.~V., {Lange-Vagle}, D., {Marchesini}, D., {et~al.} 2018, \apjs,
  235, 14

\bibitem[{{Smith} \& {Wright}(1980)}]{sw80}
{Smith}, M.~G., \& {Wright}, A.~E. 1980, MNRAS, 191, 871

\bibitem[{{Stern} {et~al.}(2005){Stern}, {Eisenhardt}, {Gorjian}, {Kochanek},
  {Caldwell}, {Eisenstein}, {Brodwin}, {Brown}, {Cool}, {Dey}, {Green},
  {Jannuzi}, {Murray}, {Pahre}, \& {Willner}}]{stern05}
{Stern}, D., {Eisenhardt}, P., {Gorjian}, V., {et~al.} 2005, ApJ, 631, 163

\bibitem[{{Treu} {et~al.}(2016){Treu}, {Schmidt}, {Brammer}, {Vulcani}, {Wang},
  {Bradac}, {Dijkstra}, {Dressler}, {Fontana}, {Gavazzi}, {Henry}, {Hoag},
  {Huang}, {Jones}, {Kelly}, {Malkan}, {Mason}, {Pentericci}, {Poggianti},
  {Stiavelli}, {Trenti}, \& {von der Linden}}]{treu16}
{Treu}, T., {Schmidt}, K.~B., {Brammer}, G.~B., {et~al.} 2016, VizieR Online
  Data Catalog, J/ApJ/812/114

\bibitem[{{Vanden Berk} {et~al.}(2004){Vanden Berk}, {Wilhite}, {Kron},
  {Anderson}, {Brunner}, {Hall}, {Ivezi{\'c}}, {Richards}, {Schneider}, {York},
  {Brinkmann}, {Lamb}, {Nichol}, \& {Schlegel}}]{vb04}
{Vanden Berk}, D.~E., {Wilhite}, B.~C., {Kron}, R.~G., {et~al.} 2004, \apj,
  601, 692

\bibitem[{{Veilleux} \& {Osterbrock}(1987)}]{vo87}
{Veilleux}, S., \& {Osterbrock}, D.~E. 1987, ApJS, 63, 295

\bibitem[{{Vestergaard} \& {Peterson}(2006)}]{VP06}
{Vestergaard}, M., \& {Peterson}, B.~M. 2006, \apj, 641, 689

\bibitem[{{Villforth} {et~al.}(2010){Villforth}, {Koekemoer}, \&
  {Grogin}}]{villforth10}
{Villforth}, C., {Koekemoer}, A.~M., \& {Grogin}, N.~A. 2010, \apj, 723, 737

\bibitem[{{Villforth} {et~al.}(2012){Villforth}, {Sarajedini}, \&
  {Koekemoer}}]{villforth12}
{Villforth}, C., {Sarajedini}, V., \& {Koekemoer}, A. 2012, \mnras, 426, 360

\bibitem[{{Zuo} {et~al.}(2012){Zuo}, {Wu}, {Liu}, \& {Jiao}}]{zuo12}
{Zuo}, W., {Wu}, X.-B., {Liu}, Y.-Q., \& {Jiao}, C.-L. 2012, \apj, 758, 104

\end{thebibliography}
